\documentclass[a4paper,11pt]{article}
\usepackage{jheppub_mod} 

\usepackage{comment}

\usepackage{graphicx}
\usepackage{subcaption}
\usepackage{setspace} 
\usepackage[utf8]{inputenc}
\usepackage{amsmath,amssymb,amsfonts,bm}
\usepackage{graphicx}
\usepackage{multirow}
\usepackage{booktabs,tabulary}
\usepackage[usenames,dvipsnames]{color}
\usepackage{siunitx}
\allowdisplaybreaks
\usepackage{bbm}
\usepackage{pgf}
\usepackage{lmodern}
\usepackage{import}
\usepackage{float}

\usepackage{orcidlink}

\newcommand{\Omnes}{Omn\`{e}s }
\newcommand{\bra}[1]{\left\langle #1 \right|}
\newcommand{\ket}[1]{\left| #1 \right\rangle}
\newcommand{\Disc}{\textrm{Disc}}
\newcommand{\Mcal}{\ensuremath{\mathcal{M}}} 
\newcommand{\Acal}{\ensuremath{\mathcal{A}}} %
\newcommand{\Bcal}{\ensuremath{\mathcal{B}}} %
\newcommand{\diff}{\mathrm{d}}
\newcommand{\ACP}{\mathcal{A}_{\text{CP}}}
\newcommand{\GeV}{\,\text{GeV}}

\renewcommand{\vec}{\bm}

\newcommand{\bonn}{Helmholtz-Institut f\"ur Strahlen- und Kernphysik, Bethe Center for Theoretical Physics, and Cluster of Excellence  ``Color meets Flavor'', Universit\"at Bonn, 53115 Bonn, Germany}

\newcommand{\ucm}{Departamento de Física Teórica and IPARCOS, Universidad Complutense de Madrid, 28040, Madrid, Spain}

\newcommand{\camp}{Departamento de Raios Cósmicos e Cronologia, Universidade Estadual de Campinas, 13083-860, Campinas, Brazil}

\newcommand{\fzj}{Institute for Advanced Simulation and Cluster of Excellence  ``Color meets Flavor'', Forschungszentrum J\"ulich, 52425 J\"ulich, Germany}

\newcommand{\usi}{Center of Particle Physics (CPPS) and Cluster of Excellence  ``Color meets Flavor'', Theoretische Physik 1, Universit\"at Siegen, 57068 Siegen, Germany}


\title{\boldmath A dispersive method to study CP asymmetries in hadronic multi-body \boldmath{$B$} decays }

\author[a,1]{L.~A.~Heuser,\note{These authors contributed equally to this work.}}
\author[b,1]{A.~Reyes-Torrecilla,}
\author[c]{C.~Hanhart,}
\author[a]{Y.~Huang,}
\author[a]{B.~Kubis,}
\author[d]{P.~C.~Magalh\~aes,}
\author[e]{T.~Mannel,}
\author[b,a]{ and J.~R.~Pel\'aez}

\affiliation[a]{\bonn}
\affiliation[b]{\ucm}
\affiliation[c]{\fzj}
\affiliation[d]{\camp}
\affiliation[e]{\usi}

\emailAdd{heuser@hiskp.uni-bonn.de, albrey01@ucm.es, c.hanhart@fz-juelich.de, huang@hiskp.uni-bonn.de, kubis@hiskp.uni-bonn.de, pcmaga@unicamp.br, mannel@physik.uni-siegen.de, jrpelaez@ucm.es}

\abstract{We present the details of a dispersive method to construct the amplitudes for hadronic multibody $B$ decays based on the universality of pairwise hadronic final-state interactions at low invariant masses. This approach allows us to
split the amplitude into source terms controlled by phenomenological parameters 
and final-state interactions, governed
by the precise knowledge of two-body dynamics, both resonant and non-resonant. As a concrete application, we make use of the well-determined low-energy pion--pion ($\pi\pi$) interactions to understand the enhanced localized CP violation observed in $B^{\pm}\rightarrow K^{\pm}\pi^+\pi^-$. Fitting only angle-integrated CP-asymmetry data, the method successfully predicts the Dalitz plot differential distribution of events and of the CP asymmetry in the low-energy region.  This allows for a better understanding of the large localized CP asymmetry recently reported by LHCb, which is shown not to be  driven by an absolute enhancement of CP-violating effects, but by a suppression of the CP-conserving part resulting from the interplay of several partial waves. Moreover, we show that the widely-neglected non-resonant isospin-2 contributions play an essential role in the description of CP violation in this system. The method can be straightforwardly adapted to other multibody decays, where final-state two-hadron interactions drive the CP violation.}

\begin{document}
\maketitle
\flushbottom

\section{Introduction}
\label{sec:intro}

Three-body non-leptonic decays of heavy mesons provide a rich environment for studying strong-interaction dynamics; however, they are also notoriously difficult to describe theoretically. 
Although QCD factorization (QCDF) has achieved considerable success in describing two-body decays, the theoretical treatment of three-body decays is still far from complete~\cite{Krankl:2015fha}. 

In the decays of bottom hadrons into non-charmed final states, the Standard Model (SM) predicts 
sizable CP asymmetries, originating from the interplay between the weak phase encoded in the 
Cabibbo--Kobayashi--Maskawa (CKM) matrix~\cite{Cabibbo:1963yz,Kobayashi:1973fv} and the strong 
phases associated with hadronic matrix elements. Unlike in two-body decays, in three-body decays, the strong phases depend
on two kinematic variables, resulting in a non-trivial Dalitz distribution of CP asymmetries, 
mapping out the strong phases of the amplitudes. 
This has been confirmed by LHCb~\cite{Aaij:2013sfa,Aaij:2013bla,Aaij:2014iva} for $B^\pm\to h_1^\pm h_2^+ h_2^-$ decays, where $h_n^\pm=\pi^\pm,K^\pm$.

LHCb reported integrated CP asymmetries of the order of a few percent, in accordance with SM expectations.  
However, in their most recent analysis with an integrated
luminosity of $5.9\,\text{fb}^{-1}$~\cite{LHCb:2022fpg}, they find  
much larger asymmetries in localized regions of the  Dalitz plot, which are connected 
to hadronic final-state interactions (FSI), especially in the $S$~wave. These can be qualitatively 
understood by employing simple resonance models~\cite{Aaij:2019hzr,Aaij:2019jaq,Aaij:2019qps}. 
For example, it was found that the $\pi\pi\rightarrow K\bar K$  rescattering in $B^\pm\rightarrow \pi^\pm K^+K^-$leads to the largest CP asymmetry reported so far, with  $(-66\pm 4 \pm 2)\%$~\cite{Aaij:2019qps}. The prominent role of the $S$-wave $\pi\pi \leftrightarrow K\bar K$ rescattering mechanism has also been identified as a key ingredient in understanding the opposite signs of the integrated CP asymmetries measured in $B^\pm \to \pi^\pm\pi^+\pi^-$ and $B^\pm\to \pi^\pm K^+ K^-$ decays in the $1$ to $1.5 \GeV$ region of the $h_2^+ h_2^-$ invariant mass~\cite{Bhattacharya:2013cvn,Bediaga:2013ela,AlvarengaNogueira:2015wpj,Garrote:2022uub}.

As far as the theoretical description is concerned, the rich structure of the Dalitz-CP-asymmetry distribution of bottom-hadron decays cannot be properly reproduced within the QCDF
approach, which is formally an expansion in $1/m_B$ where the power-suppressed contributions are 
not known. Clearly, the details of the structures are related to such power corrections, and hence 
an approach that treats the hadronic part of the decay systematically is of utmost importance.

Usually in amplitude analyses of the decays of $B$ mesons into three light hadrons~\cite{Aaij:2019qps,Aaij:2019jaq}, resonances are parametrized by 
Breit--Wigner functions or variants thereof. Such descriptions are intrinsically model- and channel-dependent and, especially when several resonances contribute to the same partial wave
or when complex couplings are used, violate unitarity. Moreover, they cannot incorporate the constraints implied by chiral symmetry, fail to describe the rich structure of $\pi\pi$ scattering including the non-Breit--Wigner-like $f_0(500)$ and $f_0(980)$ resonances, and usually neglect non-resonant partial-wave contributions. 
A number of recent theoretical works aimed at improving the description of charmless three-body $B$ decays~\cite{Furman:2005xp,El-Bennich:2006rcn,El-Bennich:2009gqk,Cheng:2016shb,Boito:2017jav,Cheng:2020ipp} (cf.\ also the reviews Refs.~\cite{Bediaga:2020qxg,Chen:2021ftn}).  
Most of these studies are based on a factorization of the hadronic matrix elements into a heavy-to-light transition form factor and a timelike light-meson form factor. The latter encodes the effects of soft rescattering and was typically modeled through phenomenological descriptions incorporating a limited set of $S0$-, $P$-, and $D0$-wave resonances, such as isobar models, Breit--Wigner parametrizations with background contributions, $K$-matrix approaches, or more sophisticated
form factor formulations. In those analyses, non-resonant $\pi\pi$ or $K\pi$ partial waves with isospin $I>1$ are generally neglected.

\begin{sloppypar}
Recently, we introduced a new method to analyze reactions of the type $B^\pm\to h_1^\pm h_2^+h_2^-$~\cite{Heuser:2025mnk}. 
In that work, we improved the description of hadronic final-state interactions by treating them in a model-independent manner through dispersion theory. This framework is, by construction, consistent with two-body unitarity and allows for a direct implementation of the high-precision phase-shift information available for the two-pion system~\cite{Ananthanarayan:2000ht,Colangelo:2001df,Garcia-Martin:2011iqs,Caprini:2011ky,Pelaez:2024uav}. As a result, the production amplitudes inherit the correct phase motion in each partial wave, which in the elastic regime is identical to the scattering phase: a key ingredient for a reliable description of CP-violating observables. The method was implemented in the elastic $\pi\pi$ scattering domain, specifically $m_{\pi\pi}<2M_K\sim 1\GeV$, a range that captures most of the sizable localized CP asymmetry measured in $B^\pm\to K^\pm\pi^+\pi^-$. 
In this paper, we provide more technical details of the formalism, additional results that allow us to better understand the origin of the mentioned large CP violation, and explain the analysis procedure in detail.
\end{sloppypar}

\begin{figure}[t]
\begin{subfigure}[t]{0.45\textwidth}
  \includegraphics[width=\linewidth]{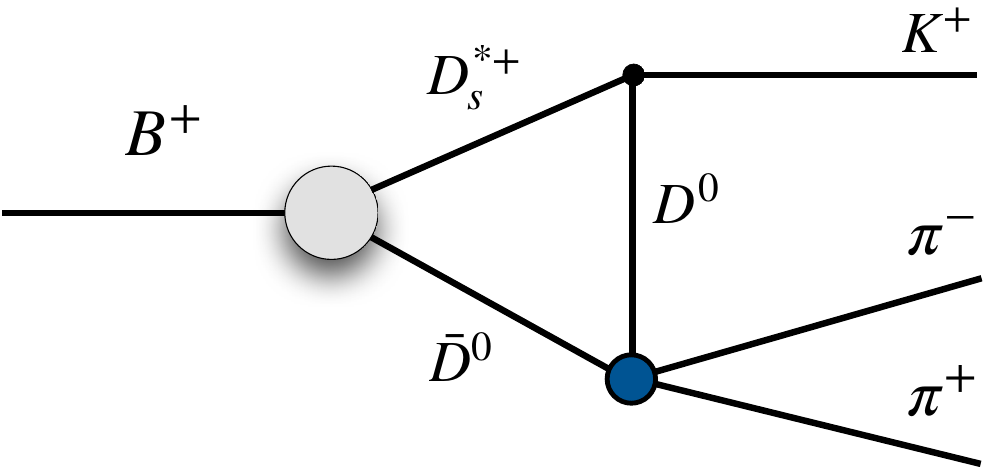} 
   \caption{}
   \label{fig:sub1}
\end{subfigure}
\hfill 
\begin{subfigure}[t]{0.47\textwidth}
   \includegraphics[width=\linewidth]{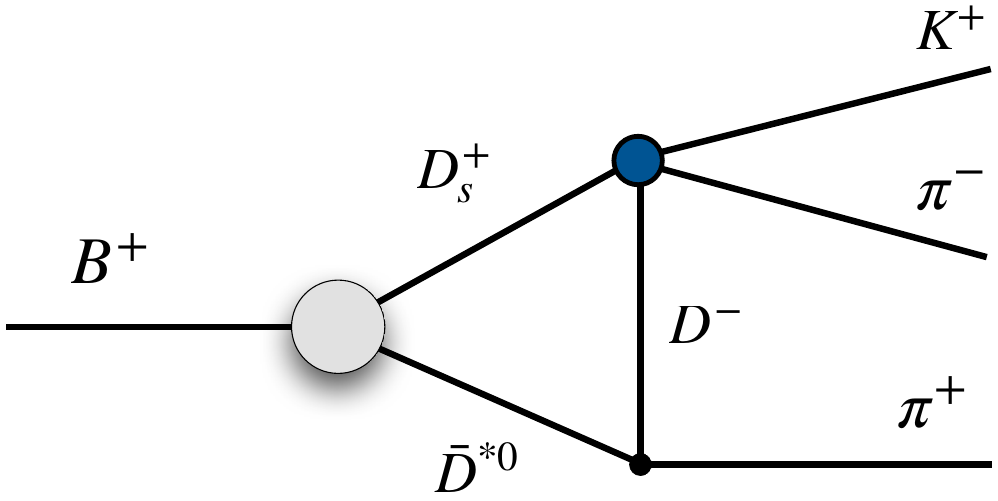}
   \caption{}
   \label{fig:sub2}
\end{subfigure}
\caption{Typical topologies of the $c\bar c$ loop at the meson level whose CP-even imaginary parts we subsume in the sources. 
In addition to cuts in $M_B^2$, diagram (b) possesses a left-hand cut with respect to the pion--pion invariant mass.  We neglect the energy dependence of these imaginary parts as explained in the main text, and approximate them by constants.}
\label{fig:triangle}
\end{figure}

It should be stressed that we do not treat the three-hadron final state completely, but build the formalism on three assumptions/approximations:
\begin{enumerate}
    \item  The weak interaction produces hadrons on a small distance scale, forming a ``source'' for outgoing hadrons, constructed in the spirit of Refs.~\cite{Aitchison:1972ay,Chung:1995dx,Gardner:2001gc,Meissner:2000bc,Daub:2015xja}.
    We assume that those sources show only a weak $m_{\pi\pi}$ dependence that can be parametrized by first-order polynomials.

\item  For small $m_{\pi\pi}$, the pion pair has a large recoil against the $BK$ system, of the order of the $B$-meson mass, $M_B$. 
Hence, the $K\pi$ invariant mass is $\mathcal{O}(M_B)$, and the kaon final-state rescattering is assumed negligible. 

\item 
In addition to cuts in the invariant mass of the two-body subsystem, partial waves of three-body decay amplitudes typically have left-hand cuts due to crossed-channel dynamics.
Furthermore, beyond cuts in two-particle variables, they also have imaginary parts in the dependence on the mass of the decaying particle (here: $M_B$), due to the mere fact that it is heavy enough to decay.
In the reactions studied here, both imaginary parts in the decay mass and left-hand cuts still emerge even when final-state rescattering with the large-recoil kaon is neglected, in particular via intermediate open-charm-meson pairs; cf.\ Fig.~\ref{fig:triangle} for sample diagrams.  These are quantitatively relevant, as the decay into open-charm states is strongly CKM favored.
As an example for a diagram containing a left-hand cut, consider Fig.~\ref{fig:triangle}(b).
In the kinematics studied in this work ($m_{\pi\pi}<1$~GeV), the $D_s^+D^-$ intermediate state can go on-shell and generate an $m_{\pi\pi}$ dependent imaginary part; in contrast to this, the $D^0\bar D^0$ intermediate state of diagram (a) of the same figure cannot go on-shell in the considered mass range of small $m_{\pi\pi}$.
Both diagrams (a) and (b) contain cuts due to $M_B > M_{D_s^*}+M_{D^0}$ and $M_B > M_{D_s}+M_{D^{0*}}$, respectively. 
In this study, we do not treat such reaction chains dynamically, but assume that their non-trivial effects can be parametrized by CP-even complex numbers, again assumed to be only slowly varying in the energy range considered and therefore approximated by constants.
This assumption appears natural, since $2M_D\gg 1$~GeV.
\end{enumerate}
Under these assumptions, all strong energy dependencies are driven
by the $\pi\pi$ FSI, which are universal and a long-distance effect.

This article is structured as follows.
In Sec.~\ref{sec:FSI}, we summarize our approach to a dispersion-theoretical description of low-energy pion--pion final-state interactions.
In Sec.~\ref{sec:sources}, we discuss the structure of the source terms we assume for the production of the pion pairs, and how they are related to the weak-interaction transition operators.
The definitions of the observables studied in the application to charmless three-body $B$-meson decays are given in Sec.~\ref{sec:applicationtoB3M},
and fits to data are presented in Sec.~\ref{sec:Fittodata}.
As only angle-integrated data are fitted, we can postdict the corresponding Dalitz-plot distributions in Sec.~\ref{sec:Dalitzplot}, 
and dissect our results for an explanation of large localized CP violation in Sec.~\ref{sec:largelocalCP}; in particular, the important role of the small, non-resonant $S2$ partial wave is discussed in Sec.~\ref{sec:S2effect}.
A more detailed comparison to the existing literature on charmless hadronic three-body decays of $B$ mesons, with particular focus on the production operators, is provided in Sec.~\ref{sec:literature}.
We summarize our findings in Sec.~\ref{sec:summary}.
Some additional background information and more technical details are relegated to the appendices.

\section{Universal final-state interaction and Omn\`es formalism}\label{sec:FSI}

While the formalism presented below is rather universal in principle, 
to ease the presentation in the following, we focus on the specific decays 
\begin{equation}
B^\pm(p_B)\to K^\pm(p_K) \pi^+(p_+)\pi^-(p_-) \,.
\end{equation}
The reaction can be fully characterized by the Mandelstam variables 
\begin{equation}
\begin{split}
&s=(p_+ + p_-)^2\equiv m_{\pi\pi}^2 \,,
\\&t=(p_K+p_{\mp})^2=(p_B-p_{\pm})^2\equiv m_{K\pi}^2 \,.
\end{split}
\end{equation}
In the following, we will work with partial-wave-decomposed amplitudes, which only leave non-trivial dynamical dependence on $s$.  

We restrict ourselves to $\pi\pi$ invariant masses below $1\GeV$.
In this kinematic regime, the partial waves of relevance are the isoscalar $S0$ wave, including the resonances $f_0(500)$ and $f_0(980)$ (with the latter known to strongly couple $\pi\pi$ and $K\bar K$ channels), the isospin-2 $S2$ wave free of resonances, and the isovector $P$ wave with the prominent contribution from the $\rho(770)$. The potential influence of the isoscalar $D0$~wave, which includes the $f_2(1270)$ resonance, is discussed in Appendix~\ref{app:fitparams}. Its effect is irrelevant at such invariant masses, and thus we have neglected $D$ and higher partial waves. However, the $f_0(980)$ lies practically at $K\bar{K}$ threshold and couples strongly to two kaons. Therefore, it is convenient to use a $\pi\pi$--$K\bar{K}$ coupled-channel formalism, even though we are still in an elastic region. As we will discuss below, the additional freedom entailed in this coupled-channel treatment, corresponding to varying relative strengths in the production of pion and kaon pairs, can be represented via the flavor structure of the isoscalar sources, in a way that allows us to establish links to the weak transition operators.
As a result, we will split the $S0$ wave according to
\begin{equation}
\Acal_{S0}^{\pm}=\Acal_{S0n}^{\pm}+\Acal_{S0s}^{\pm} \, ,
\end{equation}
 where each summand is connected to operator structures of the non-strange-quark and strange-quark source, respectively.

The decay amplitude can be written as
\begin{equation}
    \Acal^{\pm}(s,t)=\sum_i f_i(s,t)\Acal_i^{\pm}(s),
\end{equation}
with $i\in\{S0n, S0s, S2, P\}$. The helicity angle $\theta$ between $K^{\pm}$ and $\pi^{\mp}$, in the  $\pi^+\pi^-$ center-of-mass frame, is carried by $f_i(s,t)$
\begin{equation}
    f_i(s,t) = \left\lbrace\begin{array}{rl}
&1 \ \ \mbox{for} \ i \in \{S0n,S0s,S2\} , \\[1mm]
&t-u=g(s)z(s,t) \ \ \mbox{for} \ i = P,
\end{array}
\right.
\end{equation}
and we have defined 
\begin{equation}
    z(s,t)=\cos\theta, \qquad
        g(s)=-\sigma_{\pi}(s)\lambda^{1/2}(s,M_K^2,M_B^2),\qquad
        \sigma_h(s)=\sqrt{1-\frac{4M_h^2}{s}},
\end{equation}
where $\lambda(x,y,z) = x^2+y^2+z^2-2xy-2yz-2xz$ is the K\"all\'en function.

 Our dispersive analysis starts from the discontinuity relation for the production amplitude of the soft pion pair. For $i=S2, P$, this reads, in partial-wave projected form, 
\begin{equation} 
  \Disc \, \Acal_i^{\pm}(s) = 2i \Mcal^*_i(s)\rho_{\pi}(s)\Acal_i^{\pm}(s),
   \label{resonances:eq:discAmain}
\end{equation}
with $\rho_h(s) = \sigma_h(s)/16\pi$
and where $\Mcal_i(s)$ represents the $\pi\pi$ partial-wave amplitude, which, being strictly given by the strong interaction and therefore CP-invariant, does not carry the $\pm$ superscript.
This discontinuity only appears along the real axis above the $\pi\pi$ threshold.
For purely elastic interactions, $\Mcal_i(s)$ can be parametrized in terms of a phase shift $\delta_i(s)$, 
\begin{equation}
\Mcal_i(s)= \frac{e^{i\delta_i(s)}\sin\delta_i(s)}{\rho_{\pi}(s)}
\end{equation}
(which fixes our convention for the $\pi\pi$ partial waves),
such that Eq.~\eqref{resonances:eq:discAmain} simplifies to
\begin{equation}  
  \Disc \, \Acal_i^{\pm}(s) 
   = 2i \, e^{-i\delta_i(s)}\sin\delta_i(s)\Acal_i^{\pm}(s)  .
   \label{eq:discAelast}
\end{equation}
This equation admits a closed-form solution, namely~\cite{Omnes:1958hv} 
\begin{equation}
\Acal_i^{\pm}(s)=P_i(s)\Omega_i(s)\bar{\Acal}_i^{\pm},
\label{eq:Aelast}
\end{equation}
with the \Omnes function 
\begin{equation}
    \Omega_i(s)=\exp\left\{\frac{s}{\pi}\int_{4M_\pi^2}^\infty\diff s' \frac{\delta_i(s')}{s'(s'-s-i\epsilon)}\right\}, \qquad \Omega_i(0)=1,
\end{equation} 
where $\bar{\Acal}_i^{\pm}$ are the constants that parametrize the source, and $P_i(s)$ is a function free of right-hand cuts up to inelastic thresholds, for simplicity to be parametrized by a polynomial, and normalized according to $P_i(0)=1$. 
This form of $\Acal_i^{\pm}(s)$ reproduces Watson's theorem~\cite{Watson:1954uc}, which states that the phase of the
production amplitude agrees with that of the scattering amplitude for elastic partial waves.
In cases where there are no prominent left-hand cuts 
and the energy dependence induced by crossed channels can be neglected, the polynomial $P_i(s)$ in Eq.~\eqref{eq:Aelast}
is often well approximated by just a constant (see Ref.~\cite{Daub:2015xja} for a study
for $\bar{B}_{d,s}\to J/\psi \pi^+\pi^-$).
Whereas the \Omnes function describes the universal hadron--hadron interactions, the $P_i(s)$ are reaction- and partial-wave-specific, encoding the uncertainties related to the high-energy extension of the phases, and some crossed-channel effects~\cite{Kubis:2015sga}.

For the isoscalar scalar channel, or $S0$ wave, the analysis is more involved, as the proximity of the $K\bar K$ threshold necessitates a coupled-channel framework. In this case, we proceed as follows: we write 
\begin{equation} \label{eq:M2channel}
\left(\Mcal_{ab}(s)\right) {=} \left(\! \begin{array}{cc}
\dfrac{\eta(s) e^{2 i \delta_{S0}} {-} 1}{2i {\rho}_{\pi} (s)} & |g(s)| e^{i \psi_{S0}}  \\[6pt]
|g(s)| e^{i\psi_{S0}} & \dfrac{\eta(s) e^{2 i (\psi_{S0} - \delta_{S0})} {-} 1}{2i {\rho}_K (s)}  \end{array}\! \right),
\end{equation}
where $\Mcal_{ab}(s)$ are the scalar scattering partial-wave amplitude elements between the channels $1\!=\!\pi\pi$ and $2\!=\!K\bar{K}$. They are parametrized in terms of two phase shifts, where $\psi_{S0}$ is the $\pi\pi\to K\bar K$ phase shift in the $S0$ channel (whose dependence on $s$ has been omitted for brevity), along with the elasticity
\begin{equation}
\eta(s) = \sqrt{1-4\rho_{\pi}(s) \rho_K(s)|g(s)|^2\Theta(s-4M_K^2)},
\end{equation}
which is related to the modulus $|g(s)|$ of the corresponding $\pi\pi\to K\bar K$ partial wave~\cite{Cohen:1980cq,Etkin:1981sg,Buettiker:2003pp,Pelaez:2018qny,Pelaez:2020gnd}.
Therefore, the discontinuity equation \eqref{resonances:eq:discAmain} now contains two terms, accounting for $\pi\pi \rightarrow K\bar{K}$ rescattering:
\begin{equation}  
\begin{split}
  \Disc \, \Acal_{i}^{\pm} &= 2i \left(\Mcal_{11}^*\rho_{\pi}\Acal^{\pm}_{i}+\Mcal^*_{12}\rho_{K}\Bcal^{\pm}_{i}\right),\\
  \Disc \, \Bcal_{i}^{\pm} &= 2i \left(\Mcal_{21}^*\rho_{\pi}\Acal^{\pm}_{i}+\Mcal^*_{22}\rho_{K}\Bcal^{\pm}_{i}\right),
\end{split}
   \label{resonances:eq:coupleddiscAmain}
\end{equation}
where $\Bcal_i^{\pm}$ is the corresponding $K\bar{K}$ production amplitude, $i\in\{S0n,S0s\}$, and, again, the $s$ dependence has been suppressed. 
As before, these discontinuities only appear along the real axis above the respective thresholds.
There is no closed-form solution for the resulting \Omnes matrix. It has to be calculated numerically by solving the integral equation~\cite{Donoghue:1990xh,Moussallam:1999aq,Descotes-Genon:2000byu,Hoferichter:2012wf,Daub:2012mu,Celis:2013xja,Daub:2015xja,Winkler:2018qyg,Ropertz:2018stk}
\begin{equation}
\Omega(s) = \frac{1}{\pi} \int_{4M_{\pi}^2}^{\infty} \frac{\Mcal^{*}(s') \Sigma (s') \Omega (s')}{s'-s-i\epsilon}\diff s',
\end{equation}
with $\Sigma(s)= \text{diag}(\rho_{\pi}(s), \rho_K(s))$ and the boundary condition $\Omega (0) = \mathbbm{1}$. 
The related production amplitude is given in terms of the \Omnes matrix by 
\begin{equation}\label{eq:2channelMOs}
\left(
\begin{array}{c}
\Acal_{i}^{\pm} (s) \\[2pt]  
\Bcal_{i}^{\pm}(s) 
\end{array}
\right)
{=}
\left(
\begin{array}{cc}
\Omega_{11} (s) & \Omega_{12} (s)\\[4pt] 
\Omega_{21}(s) & \Omega_{22} (s)
\end{array}
\right) 
\left(
\begin{array}{c}
\bar{\Acal}^{\pm}_{i} \\[2pt] 
\bar{\Bcal}^{\pm}_{i} 
\end{array}
\right).
\end{equation}
We will use the numerical solution of the \Omnes matrix from Ref.~\cite{Ropertz:2018stk}. 
Here, $\bar{\Acal}^{\pm}_{i}={\Acal}^{\pm}_{i}(0)$, $\bar{\Bcal}^{\pm}_{i}={\Bcal}^{\pm}_{i}(0)$ still denote the normalization constants at $s=0$.
We note that the simplest solution for a specific $\pi\pi$ production amplitude $\bar{\Acal}^{\pm}_{i}$ now depends on \textit{two} free parameters: apart from the overall normalization, the ratio of production strengths $\bar{\Acal}^{\pm}_{i}/\bar{\Bcal}^{\pm}_{i}$ will affect its $s$ dependence or shape. 
For an intuitive physical interpretation, it is useful to employ the two linear combinations corresponding to non-strange and strange isoscalar scalar sources and use the normalizations of the scalar form factors of pions and kaons~\cite{Donoghue:1990xh} (cf.\ the discussion in Ref.~\cite{Albaladejo:2016mad}); for simplicity, we content ourselves with the respective leading orders in the chiral expansion, such that we have
\begin{equation}
    \Omega_{S0n}\equiv\Omega_{11}+\frac{1}{2}\Omega_{12}, \qquad \Omega_{S0s}\equiv\Omega_{12}.
\end{equation}
The non-strange source ($S0n$) couples to kaons with half the strength as to pions, while a strange-quark source term ($S0s$) only contributes to pion production via a kaon--antikaon intermediate state. 
Note that Eq.~\eqref{resonances:eq:coupleddiscAmain} only allows for a common polynomial multiplying both pion and kaon production amplitudes; as a result, we will consider $P_{S0n}=P_{S0s}\equiv P_{S0}$ multiplying the whole $S0$ wave. 

\begin{figure}[t]
    \centering
    \includegraphics[width=\linewidth]{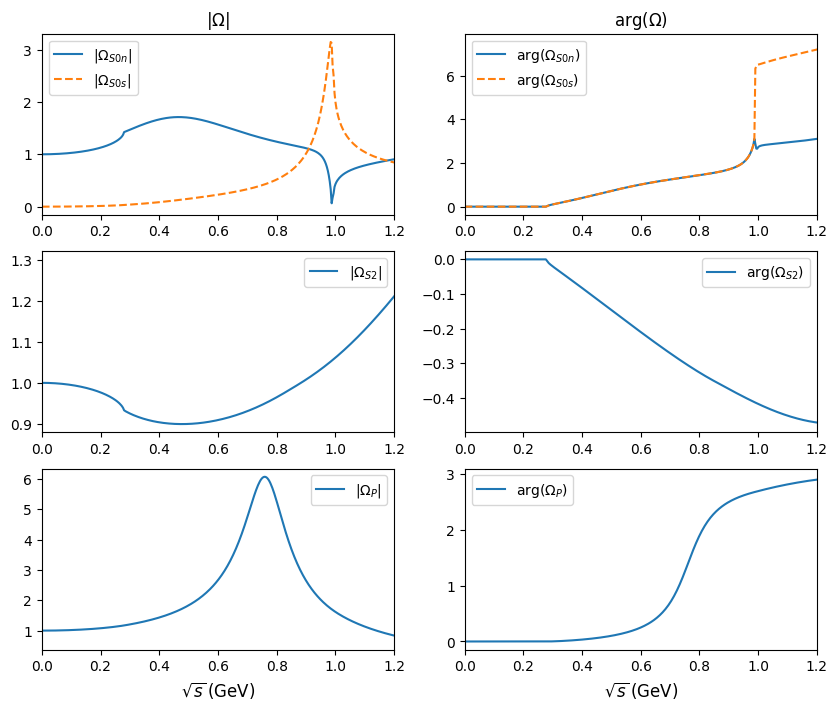}
    \caption{Moduli (left side) and phases (right side) of the Omnès functions used in this work. Note that at $s=0$, $\Omega$ is normalized to 1, except $\Omega_{S0s}(0)=0$. Watson's theorem is satisfied in the elastic region, where $\text{arg}(\Omega_{S0n})=\text{arg}(\Omega_{S0n})$.  Uncertainties are completely negligible compared to those in the source parameters or polynomial functions. Note the differences in scale between plots, particularly for the $S2$ phase. }
    \label{fig:omnes_functions}
\end{figure}

In this way, all final-state interactions are captured by \Omnes functions ($i=S2,P$) or an \Omnes matrix ($i=S0n,S0s$), and the full $B^{\pm}\rightarrow K^{\pm}\pi^+\pi^-$ amplitude parametrization will be written as 
\begin{equation}
   \Acal^{\pm}(s,t)=\sum_i f_i(s,t)P_i(s)\Omega_i(s)\bar{\Acal}_i^{\pm}. 
\end{equation}
Moduli and phases of the corresponding Omn\`es functions $\Omega_{S0n}$, $\Omega_{S0s}$, $\Omega_{S2}$, and $\Omega_P$ are displayed in Fig.~\ref{fig:omnes_functions} for illustration.
For the smooth continuation of the phases $\delta_0^2$ to $0$ and $\delta_1^1$ to $\pi$ we use the parametrization
\begin{equation}
    \delta(s)_{\rm cont}=\delta(s=\infty)-a\frac{\lambda+s_0}{\lambda+s}
\end{equation}
above the value $s=s_0$. The parameters $a$ and $\lambda$ are determined such that the phase and its derivative are smooth. 

The free parameters are the four normalization constants
$\bar{\Acal}_{S0n}$, $\bar{\Acal}_{S0s}$, and
$\bar{\Acal}_{S2}$ for the $S$~waves, 
together with $\bar{\Acal}_{P}$ for the $P$~wave,
as well as potential slope parameters in $P_i(s)$.
In the limited kinematic region described in the introduction, the $t$ dependence can be approximated polynomially and enters only via the angular dependence of the $P$~wave.

\section{Source terms}\label{sec:sources}

For any application of this method, we have to 
specify the sources. Thus, we continue by presenting the general formalism 
for the particular case of the $B \to K\pi\pi$ decay, starting with the weak transition, followed by a discussion of the contribution of the $\omega$ meson.

\subsection{The weak transition operators}

CPV in the SM is induced by the
interplay between weak and strong phases. The weak phase originates from the CKM matrix~\cite{Cabibbo:1963yz,Kobayashi:1973fv}, where in 
the standard parametrization~\cite{Chau:1984fp}, the CKM matrix element  $V_{ub}$ contains the weak phase $\gamma$, whereas $V_{cb}$, $V_{cs}$, and $V_{us}$ are taken to be real. The corresponding effective weak Hamiltonian reads
\begin{equation} \label{Heff}
H_{\rm eff} = \frac{G_F}{\sqrt{2}} \left[|V_{cb}^* V_{cs}| (\bar{b}c) (\bar{c}s) 
+ e^{i \gamma} |V_{ub}^* V_{us}| (\bar{b}u) (\bar{u}s) 
\right]+ \text{h.c.},
\end{equation}
where $(\bar{q} q')$ is the usual $V-A$ current of weak interactions and $\gamma$ is the CP-odd 
phase from the CKM matrix. 

According to our assumptions, $H_{\rm eff}$ generates the primary hadrons, which will undergo universal rescattering
from hadron--hadron interactions.
While all strong phases due to the $\pi\pi$ interaction are captured
in the \Omnes functions/matrix, there are additional sources of CP-even phases. They can originate from rescattering with the outgoing kaon, which we neglect here, and which can in principle be captured using more complicated integral equations for three-hadron final states~\cite{Khuri:1960zz,Niecknig:2015ija,Niecknig:2017ylb,Stamen:2022eda,Kou:2023kvp}.
On the other hand, they could originate from charmed-meson loops~\cite{Bediaga:2020ztp, Mannel:2020abt} as detailed below.
\begin{figure}[t]
\centering
\includegraphics[width=0.65\textwidth]{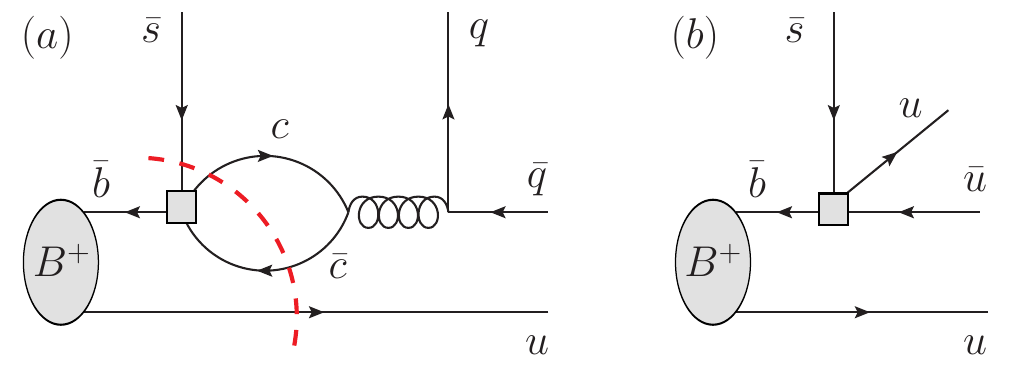}
\caption{Typical topologies for quark-level decays of $B^+\to K^+\pi^+\pi^-$ (a) with and (b) without charm loop. The gray box represents the flavor-changing weak decay process, which contains the CKM elements and CPV phase.
For $q=s$, the operator in diagram (a) provides an $\bar ss$ source,
for $q=u$ or $d$ a $\bar uu$ or $\bar dd$ source.
The red dashed line indicates the cut that generates an imaginary part.
Diagram (b) provides a $\bar uu$ source. 
}
\label{fig:ccbar}
\end{figure}
Evaluating the decay amplitude from Eq.~\eqref{Heff} involves the two matrix elements 
\begin{align}
M_{u\bar{u}} &= \langle K^+ \pi^+ \pi^- |(\bar{b}u)(\bar{u}s) | B^+ \rangle \,, \notag\\ 
    M_{c\bar{c}} &= \langle K^+ \pi^+ \pi^- |(\bar{b}c)(\bar{c}s) | B^+ \rangle \,,
\end{align}
which we parametrize using the ansatz described above. 
They differ only by the up-type quarks involved. While in $M_{u\bar{u}}$ the up quark can appear directly in the final state, the charm quarks in $M_{c\bar{c}}$ must annihilate, as illustrated in Fig.~\ref{fig:ccbar}. At the hadronic level, this implies that intermediate states such as $\bar D^{(*)} D_s^{(*)}$ can rescatter into the final $K\pi\pi$ state, illustrated in Fig.~\ref{fig:triangle}. The phases generated by the rescattering of these charmed states, represented by the $c\bar c$ cut (red line) in Fig.~\ref{fig:ccbar}(a), will thus introduce an imaginary part in the source term associated with $M_{c\bar{c}}$ for each partial wave. These phases are CP-even and originate from the short-distance dynamics governed by the charm-quark mass. That strong energy dependence occurs only near charm thresholds; in the region relevant to our analysis, such variation is not expected. Consequently, these phases can be incorporated into the source terms $\bar{\Acal}_{i}^{\pm}$ as constant imaginary contributions. In an approach like QCD factorization (see, e.g., Ref.~\cite{Klein:2017xti}),
the charm quark is included at the quark level. However, with this ansatz, the complex phenomenology of differential CP asymmetries over the Dalitz plot cannot be described~\cite{Mannel:2020abt}. For this reason, we shall fix the imaginary parts of $M_{c\bar{c}}$ by fits to the data. 

The matrix element $M_{u\bar{u}}$ contains two contributions. One is analogous to Fig.~\ref{fig:ccbar}(a), replacing the charm quark with an up quark; this diagram would contribute only to the isoscalar $q\bar q$ pair production, or the isoscalar part of the weak effective Hamiltonian.  On the other hand, the second contribution in which the up quarks propagate directly into the final state (see Fig.~\ref{fig:ccbar}(b)) will contribute to both the isoscalar and isovector parts, and we will assume the source term for each partial wave originating from this mechanism to be real.

For the reaction studied here, the contribution of $M_{c \bar{c}}$ is CKM favored, since it is multiplied by $V_{cb}V_{cs}^*$ in $H_{\rm eff}$, while $M_{u \bar{u}}$ comes with a factor $V_{ub}V_{us}^*$, so it suffers from a CKM suppression of a relative factor $\lambda^2 \sim 0.04$ in the Wolfenstein parametrization~\cite{Wolfenstein:1983yz}. However, the charm quarks appearing in 
$M_{c \bar{c}}$ have to annihilate, resulting in loop suppression. Therefore, one expects an interplay between both mechanisms.  In particular, the loop contribution to $M_{u\bar{u}}$ will be doubly small; for this reason, we effectively disregard this contribution from here on.\footnote{The influence of the $u\bar{u}$ loop was tested by including CP-violating real parts to the source parameters. However, no improvement to the fit was ovserved and the pertiment parameters were compatible with zero, which validates our assumption.}  We may thus parametrize the source term as  
\begin{equation}
\bar{\Acal}^\pm_{i} = \hat A_{i}  + e^{\pm i\gamma}\hat B_{i}  \,,
\end{equation}
where the sign on the amplitude refers to the sign of the decaying $B$ meson. 
 The $\hat A_{i}$
contain a CP-even constant imaginary part, while the 
$\hat B_{i}$ are real. 
We may, in the following, express the source terms as 
\begin{equation}
\bar{\Acal}^\pm_{i} =  a_{i} + i c_{i}  \pm i b_{i} \, , 
\label{eq:source}
\end{equation}
where
$a_{i} = {\rm Re}\, \hat A_{i} + \hat B_{i} \, \cos \gamma$,   
$c_{i} = {\rm Im}\, \hat A_{i}$, and
$b_{i} =  \hat B_{i} \, \sin \gamma$.
Dominant contributions for CP-even imaginary parts $ic_i$ arise from the typical diagram topologies in Fig.~\ref{fig:triangle}.

Since in our normalization all final-state interactions are equal to 1 at $s=0$, we have 
\begin{equation}
\Acal^\pm(0,t) =\sum_i f_i(0,t)\Acal^\pm_{i}(0)=\sum_i f_i(0,t)\bar{\Acal}^\pm_{i}  \,,
\end{equation}
which allows us to connect the source terms of Eq.~\eqref{eq:source}
to quark-level transition matrix elements, since
\begin{align}
f_i(0,t)\bar{\Acal}^\pm_{i} &= \left\langle  K^\pm [\pi^+\pi^-]_i\big|_{s=0}\right|(\bar{b}c)(\bar c s) + \text{h.c.}\left|B^\pm
\right\rangle \notag\\ & + e^{\pm i\gamma}
\left\langle  K^\pm [\pi^+\pi^-]_i\big|_{s=0}\right|(\bar b u)(\bar u s) + \text{h.c.} \left|B^\pm
\right\rangle ,
\label{eq:quarkmatele}
\end{align}
where the operators are to be understood as derived in weak effective theory, and the vertex structure was skipped for simplicity. 

It is instructive to take a closer look at the flavor structure of the source term, as it allows us to reduce the number of parameters. First of all,  $\Acal_{S0s}$ (see Eq.~\eqref{eq:2channelMOs}) in the current parametrization refers to an isoscalar scalar pion pair emerging from an $\bar ss$ source, since it originates solely from $K\bar{K} \to \pi \pi$ rescattering.
However, the CP-violating phase $\gamma$ is attached to the matrix element with the pure $\bar uu$ source. Accordingly, we need to set the parameter $b_{S0s}$ to zero.\footnote{We checked that allowing this parameter to float in the fit yields a value consistent with zero within uncertainties.} 
Moreover, we use
isospin symmetry (up to the effect of $\rho$--$\omega$
mixing discussed below) to analyze  Eq.~\eqref{eq:quarkmatele}.
The operator in the first line has zero isospin. Since the $B$
meson and the kaon both have isospin 1/2, isospin conservation tells us that the pion pair in the first matrix element can only have isospin 0 or 1.  As a result, the whole $\pi\pi$ $S2$ wave originates from the second term in Eq.~\eqref{eq:quarkmatele}, and thus we also set $c_{S2}=0$.  All in all,  there are in total 13 parameters
necessary to parametrize the production vertex of 4 partial waves, 3 each for the isoscalar scalar non-strange and isovector vector partial wave, 2 for the isoscalar scalar strange and the isotensor scalar partial wave. On top of this, we allow for non-vanishing real slope parameters in
the polynomials introduced in Eq.~\eqref{eq:Aelast}. These bring in 3 additional parameters.

\subsection[Contribution of the \texorpdfstring{$\omega$}{ω} meson]{Contribution of the \texorpdfstring{\boldmath{$\omega$}}{ω} meson}

As shown in Fig.~\ref{fig:ccbar}, diagram $(b)$ provides a
source with specified light-quark content, namely $[\bar s u\bar uu]$, while diagram $(a)$ formally provides a $[\bar s u(\bar uu+\bar dd+\bar ss)]$ source.
However, the gluon shown is necessarily
hard ($q^2\sim m_b^2$), and in this analysis we restrict ourselves to a kinematic regime 
where, in the $B$ rest frame, the $K$
recoils against the pion pair, which is restricted to small invariant masses.
Thus, it is  natural to assume that
the quark--antiquark pair from the hard gluon goes out back-to-back to be 
absorbed into the charged kaon and the source of the pion pair, respectively. 
Then, however, the chosen final state,
$K^+\pi^+\pi^-$,
only allows for $\bar uu+\bar ss$
to emerge from the hard gluon,
but not $\bar dd$, which would lead to the
production of a neutral kaon.
This logic can even be recovered in the hadronic
loops shown in Fig.~\ref{fig:triangle}: 
the hard gluon from the annihilation
of the $\bar cc$ pair present in the intermediate $D\bar D$ states needs to be
understood as being emitted in between the kaon and the pion pair.
 Thus, in this spirit diagram $(a)$ of
Fig.~\ref{fig:ccbar} provides
either a $\bar uu$ or a $\bar ss$ source
for the pion pair.

The approximate isospin symmetry of the strong interaction allows the isovector $\rho$ meson to couple to pion pairs, but strongly suppresses the respective coupling of the isoscalar $\omega$ meson. However,
as was demonstrated, e.g., in Ref.~\cite{Daub:2015xja}, the smallness of the isospin violating $\rho$--$\omega$ mixing is easily compensated in those reactions that have a significant $\omega$ production relative to the $\rho$, since the narrow $\omega$ propagator
provides an enhancement of the $\omega\to \rho$
transition of the order of $M_\omega/\Gamma_\omega\approx 90$. 
Thus we allow for the $\omega$ contribution including the known strength of the $\rho$--$\omega$ mixing~\cite{Colangelo:2018mtw,Holz:2022hwz,Colangelo:2022prz,Dias:2024zfh,Cao:2025ncx},
with the additional constraint that all coupling parameters of the $\omega$ are fixed from those of the $\rho$-channel using the observation that the source is purely of $\bar uu$-type
(the coupling of the $\omega$ to
$\bar ss$ in negligible). Following Ref.~\cite{Daub:2015xja}, we may then decompose the up-type vector current according to
\begin{equation}
    \bar u\gamma^\mu u =\frac12 (\bar u\gamma^\mu u - \bar d\gamma^\mu d) + 
    \frac12 (\bar u\gamma^\mu u + \bar d\gamma^\mu d) \ ,
\end{equation}
where the first (second) term on the right-hand side indicates the isovector (isoscalar) part of the source.
Accordingly, we conclude that the isovector and isoscalar components are fed with the same sign and strength.
It is interesting to compare this decomposition to that of the electromagnetic current:
\begin{equation}
    j_{\rm em}^\mu  =\frac12 (\bar u\gamma^\mu u - \bar d\gamma^\mu d) + 
    \frac16 (\bar u\gamma^\mu u + \bar d\gamma^\mu d) \ .
\end{equation}
Thus, in the $B^\pm\to K^\pm \pi^+\pi^-$ reaction, the effect of the $\omega$ is a factor of 3 enhanced compared to its effect on the pion electromagnetic form factor, measured in $e^+e^-\to \pi^+\pi^-$.

The reasoning provided at the beginning of this section is at most qualitative, especially since we found in this study that QCD factorization should not be used in the analysis of $B\to K\pi\pi$. However, 
we checked that releasing the $\omega$-coupling does not significantly improve the fit quality. We regard this test also as support of the assumed structure of the source term. In this sense, the $\rho$--$\omega$ mixing is acting as a probe of the production dynamics.

\section{Application to CPV in charmless three-body \textit{B} decays}
\label{sec:applicationtoB3M}

Putting everything together, we arrive at the following expression for  the differential $B^\pm\to K^\pm \pi^+\pi^-$ decay count rates: 
\begin{align} 
    \Gamma^{\pm}(s,z)&=\frac{\diff^2\Gamma^\pm}{\diff\sqrt{s}\,\diff z}(s,z)= \frac{-\sqrt{s}\,g(s)}{32 (2\pi)^3M_B^3} \vert {\cal A}^{\pm}(s,z)\vert^2 \notag\\
    &=\frac{-\sqrt{s}\,g(s)}{32 (2\pi)^3M_B^3}\sum_{i,j}f_if_j P_i P_j  
    \Big\{ \mbox{Re}\left(\Omega_i\Omega_j^*\right)(a_{i}a_{j}{+}b_{i}b_{j}{+}c_{i}c_{j})  +\mbox{Im}\left(\Omega_i\Omega_j^*\right)2a_{i}c_{j}
    \notag\\ & \hspace*{4.2cm}
    \pm\Big[\mbox{Re}\left(\Omega_i\Omega_j^*\right)2c_{i}b_{j}+\mbox{Im}\left(\Omega_i\Omega_j^*\right)2a_{i}b_{j}\Big] \Big\}, \label{eq:amplitude}
\end{align}
where, to simplify later discussions in terms of the forward and backward symmetric and asymmetric 
amplitudes, we have used $z=\cos\theta$ instead of $t$, and most of the $s$ dependence has been suppressed for brevity.
Hence, the differential decay CP-violating asymmetry reads as follows:
\begin{align}\nonumber
\Delta \Gamma_{{\rm CP}}(s,z)&\equiv\Gamma^{-}(s,z)-\Gamma^{+}(s,z)\\&=4\frac{\sqrt{s}\,g(s)}{32 (2\pi)^3M_B^3}
\sum_{i, j}f_if_j P_i P_j\Big[ a_{i}b_{j} {\rm Im}(\Omega_i\Omega_j^*)  
+ c_{i}b_{j} {\rm Re}(\Omega_i\Omega_j^*) \Big].
\end{align}
In the elastic regime, $s<1\GeV^2$, this simplifies to 
\begin{equation}
 \Delta \Gamma_{{\rm CP}}(s,z)= 4\frac{\sqrt{s}\,g(s)}{32 (2\pi)^3M_B^3}
\sum_{i,j}f_if_jP_i P_j|\Omega_i||\Omega_j|\Big[ a_{i}b_{j}\sin(\delta_i{-}\delta_j)+c_{i}b_{j}\cos(\delta_i{-}\delta_j)\Big]  ,
\label{eq:mainresult}
\end{equation}
since, as mentioned above, here the phase of the production amplitude agrees
with the scattering phase shift. Note that, below the $K\bar K$ threshold, the phases of  $\Omega_{S0n}$ and $\Omega_{S0s}$ also are the same, i.e.,
$\delta_{S0n}=\delta_{S0s}$.

In addition to the count-rate difference of the decays $B^-$ and $B^+$, the sum is an interesting observable, since it allows us to further constrain the parameters. We have, again in the elastic regime, 
\begin{align}
\Sigma \Gamma(s,z)\equiv\Gamma^{-}(s,z)+\Gamma^{+}(s,z)= &-\frac{\sqrt{s}\,g(s)}{32 (2\pi)^3M_B^3}
\sum_{i, j}f_if_j P_i P_j|\Omega_i||\Omega_j| \label{eq:mainresult2}\\
&\notag\times \Big[(a_{i}a_{j}{+}b_{i}b_{j}{+}c_{i}c_{j})\cos(\delta_i{-}\delta_j)+2a_{i}c_{j}\sin(\delta_i{-}\delta_j)\Big].
\end{align}
To increase statistics, LHCb presents its data integrated either from $z=-1$
to 0 or from 0 
to $z=1$. Since in the expressions derived above, all $z$ dependence resides in 
the $f_i$ functions, with $f_P\propto z$ being the only non-constant one, we find 
\begin{equation}
    \int_{-1}^0 \diff z f_i f_j
    = \left\lbrace\begin{array}{rl}
1 \ \  & \mbox{for} \ i,j \in \{S0n,S0s,S2\} \,, \\[1mm]
-\frac12g(s)\phantom{^2} 
 & \mbox{for} \ i  \in \{S0n,S0s,S2\}  \ \text{and} \ j=P \,,  \\[1mm]
\frac13 g(s)^2 & \mbox{for} \ i,j=P\,.
\end{array}
\right.\label{eq:angular_integration}
\end{equation}
For the other integration range, from 0 to 1, only the sign of the term in the second row changes, the $S$--$P$ interference.

LHCb provides the data separately integrated or ``projected'' over the forward ($z>0$) and backward ($z<0$) regions.
However, with the purpose of disentangling different contributions within our model, it is convenient to define the following forward--backward symmetric and antisymmetric CP differences:
\begin{align}
\Delta \Gamma_{{\rm CP}}^{(+)}(s)&=
\Delta \Gamma_{{\rm CP}}(s,z>0)+\Delta \Gamma_{{\rm CP}}(s,z<0),\label{eq:diffFBsym} \\
\Delta \Gamma_{{\rm CP}}^{(-)}(s)&= \Delta \Gamma_{{\rm CP}}(s,z>0)-\Delta \Gamma_{{\rm CP}}(s,z<0) ,
\label{eq:diffFBasym} 
\end{align}
and their corresponding CP-symmetric sums or yields
\begin{align}
\Sigma \Gamma^{(+)}(s)&=
\Sigma \Gamma(s,z>0)+\Sigma \Gamma(s,z<0),\label{eq:sumFBsym} \\
\Sigma \Gamma^{(-)}(s)&= \Sigma \Gamma(s,z>0)-\Sigma \Gamma(s,z<0) .
\label{eq:sumFBasym} 
\end{align}

For brevity, we also define ${S}0\equiv\{S0n,S0s\}$ and ${S}\equiv{S}0\cup \{S2\}$, which allows us to write
\begin{align}
    \Delta \Gamma_{{\rm CP}}^{(+)}(s) = 8\frac{\sqrt{s}\,g(s)}{32 (2\pi)^3M_B^3}\bigg\{&\label{eq:main_2}P_{S0}P_{S2}|\Omega_{S2}|\sum_{i\in S0}|\Omega_i|(a_{i}b_{S2}-a_{S2}b_{i})\sin(\delta_i{-}\delta_{S2})\\&\notag+P_{S0}\sum_{i\in S0}\sum_{j\in S}P_j|\Omega_i||\Omega_j|c_{i}b_{j}\cos(\delta_i{-}\delta_j)+P_P^2 \frac{g(s)^2}{3}|\Omega_P|^2c_{P}b_{P}\bigg\},
\end{align}
where, in particular, the typical line shape
of the $\rho$ resonance enters in this observable solely through the last term, which appears only when there are imaginary parts from the charm loops (namely $c_i$ parameters). 
In addition, we find for the other CP-violating linear combination
\begin{align}
    \Delta \Gamma_{{\rm CP}}^{(-)}(s)=4\frac{\sqrt{s}\,g(s)}{32 (2\pi)^3M_B^3}g(s)P_P|\Omega_P|\sum_{i\in S}P_i|\Omega_i|\Big[&\label{eq:main_3} (a_{i}b_{P}-a_{P}b_{i})\sin(\delta_i{-}\delta_P)\\&\notag+(c_{P}b_{i}+c_{i}b_{P})\cos(\delta_i{-}\delta_P)\Big].
\end{align}
Note that in this case, all contributions are linear in the $P$-wave amplitude, for an odd angular distribution can only be achieved from $S$--$P$ interferences.

\section{Fixing the source parameters with angle-integrated data}
\label{sec:Fittodata}

After describing our treatment of the FSI and the source terms, all that is left is to fix the source parameters of Eq.~\eqref{eq:source}.
This we do by fitting the high-statistics LHCb data~\cite{LHCb:2022fpg} (supplemental material) on $\Gamma^\pm(s)$, which is provided with uncertainties, integrated over forward and backward regions.  Details about the fit strategy, the values obtained for the fit parameters, their uncertainties, and the covariance matrix can be found in Appendix~\ref{app:fitparams}. 

\begin{figure}
\centering
\includegraphics[width=\textwidth]{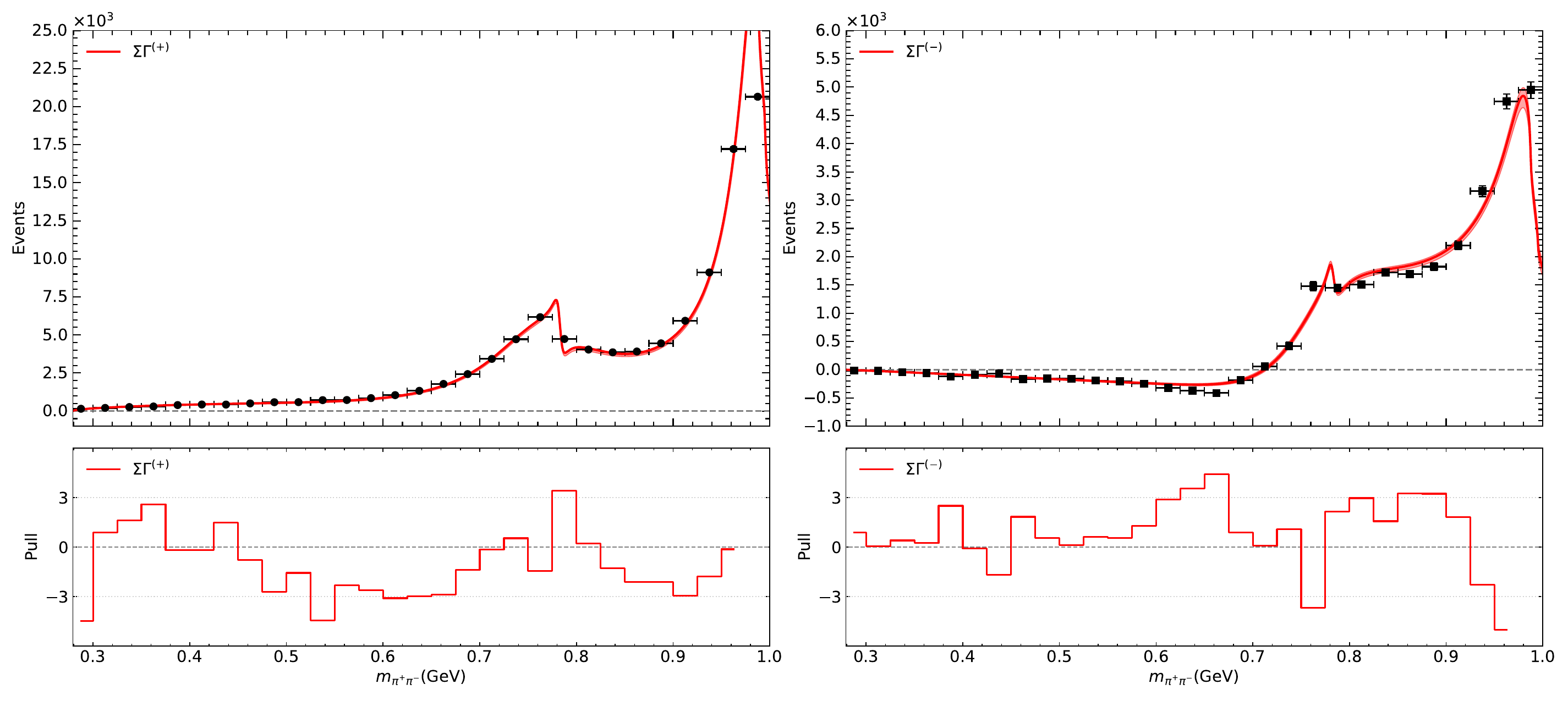}
\includegraphics[width=\textwidth]{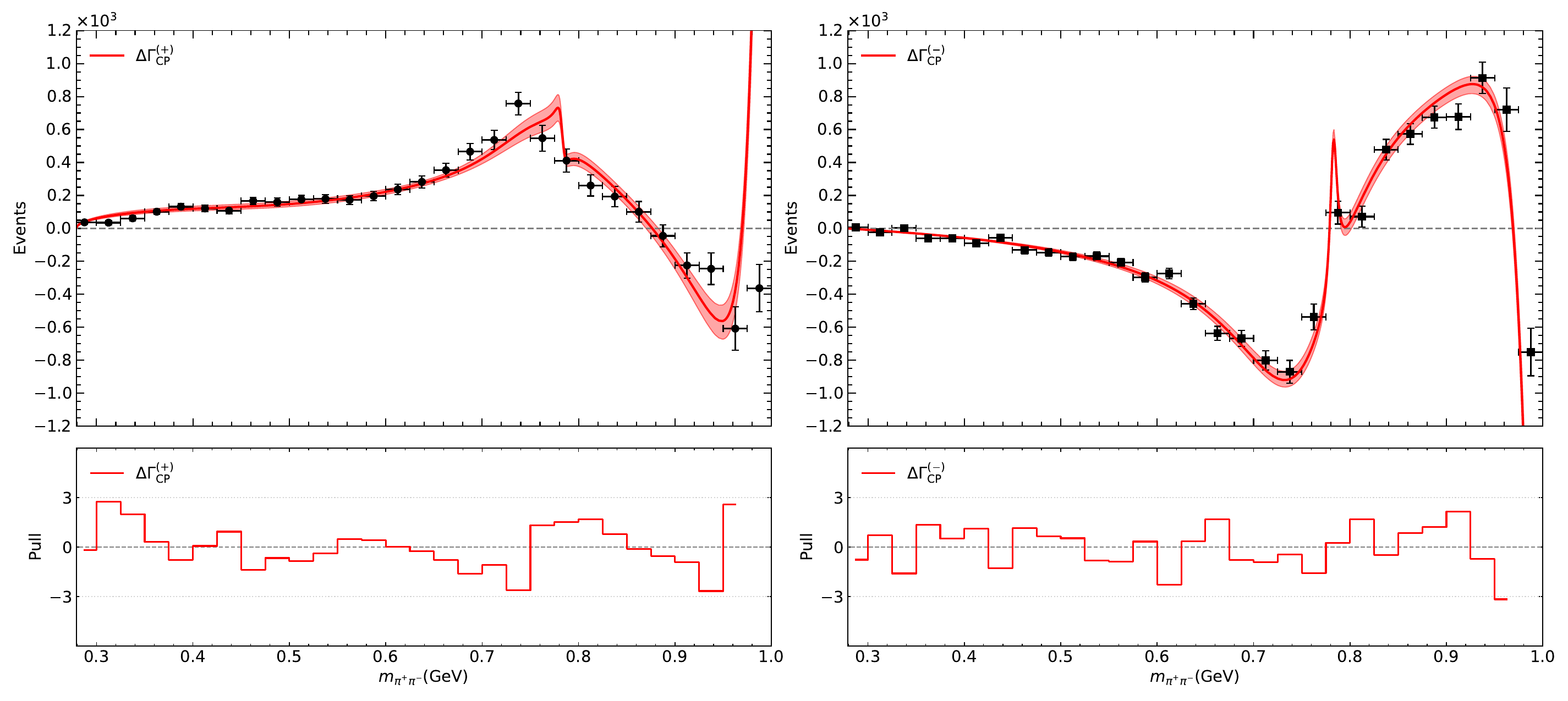}
\caption{
Results for the different distributions as functions of $\sqrt{s}$. 
The first row shows $\Sigma \Gamma^{(\pm)}$, the second row $\Delta \Gamma_{\rm CP}^{(\pm)}$. 
The first (second) column shows the forward–backward
symmetric (antisymmetric) quantities. 
Only data up to $\sqrt{s}=1\GeV$ were included in the fits.
}
\label{fig:DeltaK}
\end{figure}

The results of our fits to the four distributions $\Delta \Gamma_{\rm CP}^{(\pm)}(s)$ and $\Sigma \Gamma_{\rm CP}^{(\pm)}(s)$ are shown in Fig.~\ref{fig:DeltaK}, together with the corresponding pull distributions. These pulls are shown to facilitate comparison with the experimental data, and their magnitudes are consistent in size with those reported in LHCb analyses. Our error bands have been calculated using bootstrapping and explicitly imposing the confidence limits on the $\chi^2/\text{d.o.f.}$ as prescribed in Ref.~\cite{Press:2007ipz}.

\begin{figure}
\centering

\includegraphics[width=\textwidth]{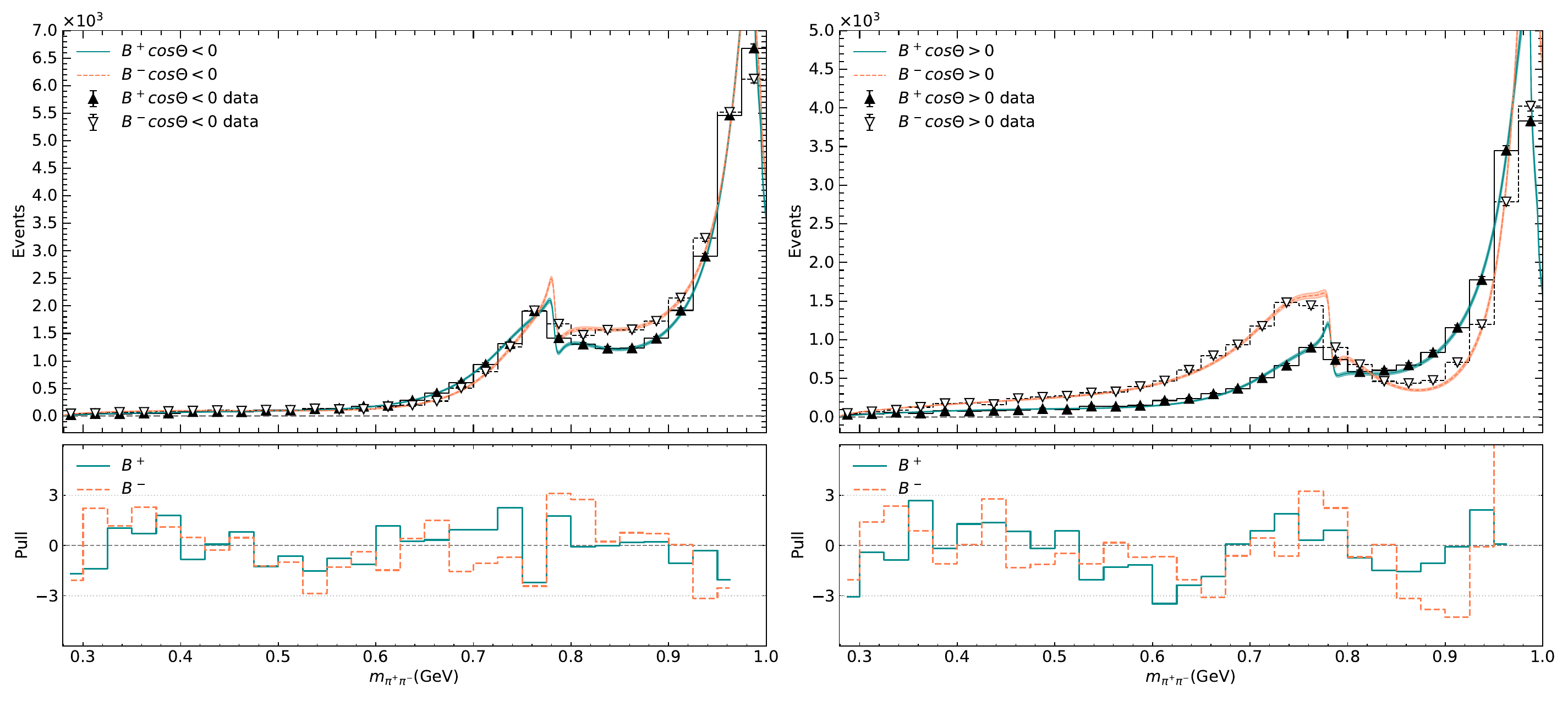}

\caption{
Results for the total yields as functions of $\sqrt{s}=m_{\pi^+\pi^-}$. The left (right) panel shows angular integration for scattering angle $\theta$ smaller (bigger) than zero for the decay widths of $B^+$ and $B^-$. Only data up to $\sqrt{s}=1\GeV$ was included in the fits.
}
\label{fig:Yields}
\end{figure}

Naturally, this same fit also provides an accurate description of the total number of events measured for both $B^+$ and $B^-$ decays, as shown in Fig.~\ref{fig:Yields}.

\begin{figure}
\centering
\includegraphics[width=\textwidth]{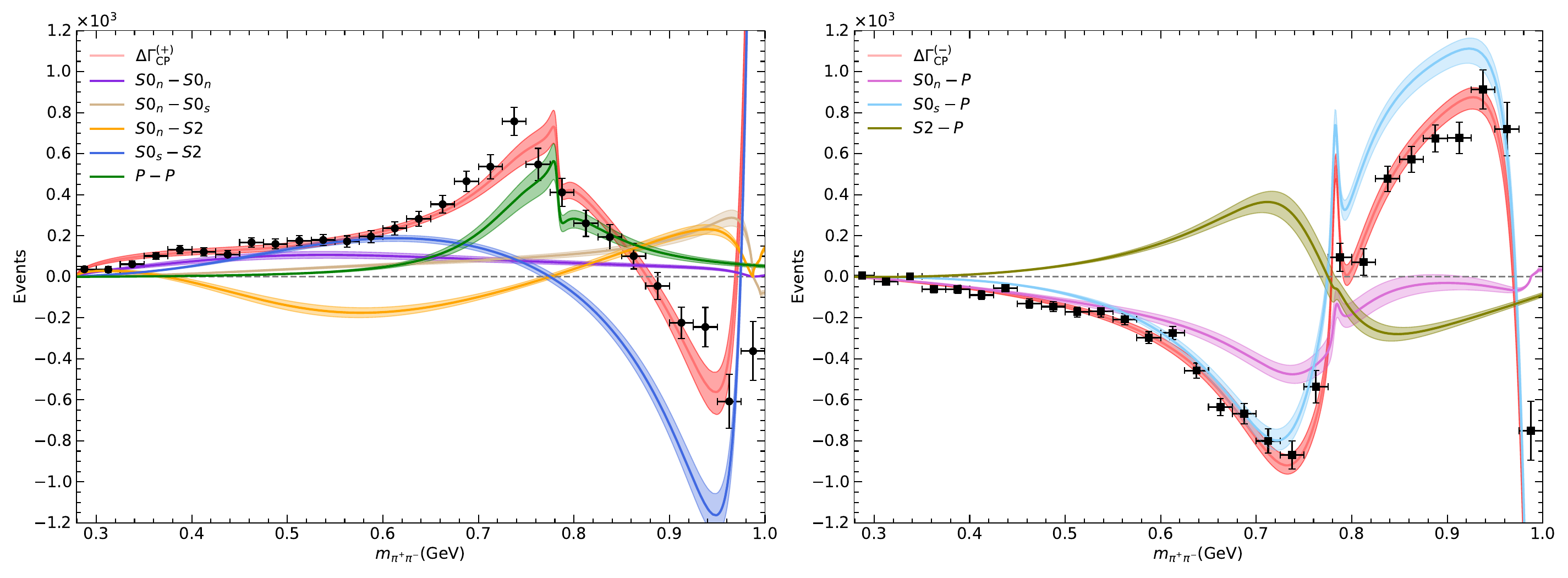}
\caption{
Results for the most important contributions to the CP differences $\Delta\Gamma^{(\pm)}_{\text{CP}}$. Note the relevance of the interference of the $S2$ wave with all other waves. 
}
\label{fig:ContributionsImportant}
\end{figure}

A key advantage of the present approach is that all parameters 
allow for physical interpretation. For instance, the coefficients $c_i$ are dominantly (favored by their CKM scaling) generated by intermediate $D^{(*)}\bar D_s^{(*)}$ pairs going on-shell in the transition $B \to D^{(*)}\bar D_s^{(*)} \to K\pi\pi$. That such a contribution is indeed required is reflected, for example, by the presence of a $\rho$ peak in $\Delta \Gamma_{{\rm CP}}^{(+)}$ in Fig.~\ref{fig:ContributionsImportant}, which according to Eq.~\eqref{eq:main_2} scales as $c_P b_P$. Moreover, as already remarked in Ref.~\cite{Heuser:2025mnk}, both CP-odd distributions require the inclusion of the isotensor $S2$~wave in order to achieve an acceptable description: in $\Delta \Gamma^{(+)}_{\text{CP}}$ it interferes with both the strange and the non-strange isoscalar $S$~waves, while in $\Delta \Gamma^{(-)}_{\text{CP}}$ it interferes with the $\rho$ amplitude. The $S2$-wave contribution was not included in previous analyses, most likely because it does not contain any resonant structure.
In Sec.~\ref{sec:S2effect} below, we will discuss its effect on the differential distributions in detail.

We regard the significant
rise of the $S0n$--$S2$ interference in $\Delta \Gamma^{(+)}_{\text{CP}}$ beyond $s=1\GeV^2$ as
non-physical, since it is driven by the sizable linear term in the polynomial multiplying
$\Omega_{S2}$ (cf.\ Eq.~\eqref{eq:Aelast}), which is typically generated
from left-hand cuts neglected here. Thus, this rise should be tamed in
an improved analysis, which, however, goes beyond the goals of this study.

\begin{figure}[t]
\centering

\includegraphics[width=\textwidth]{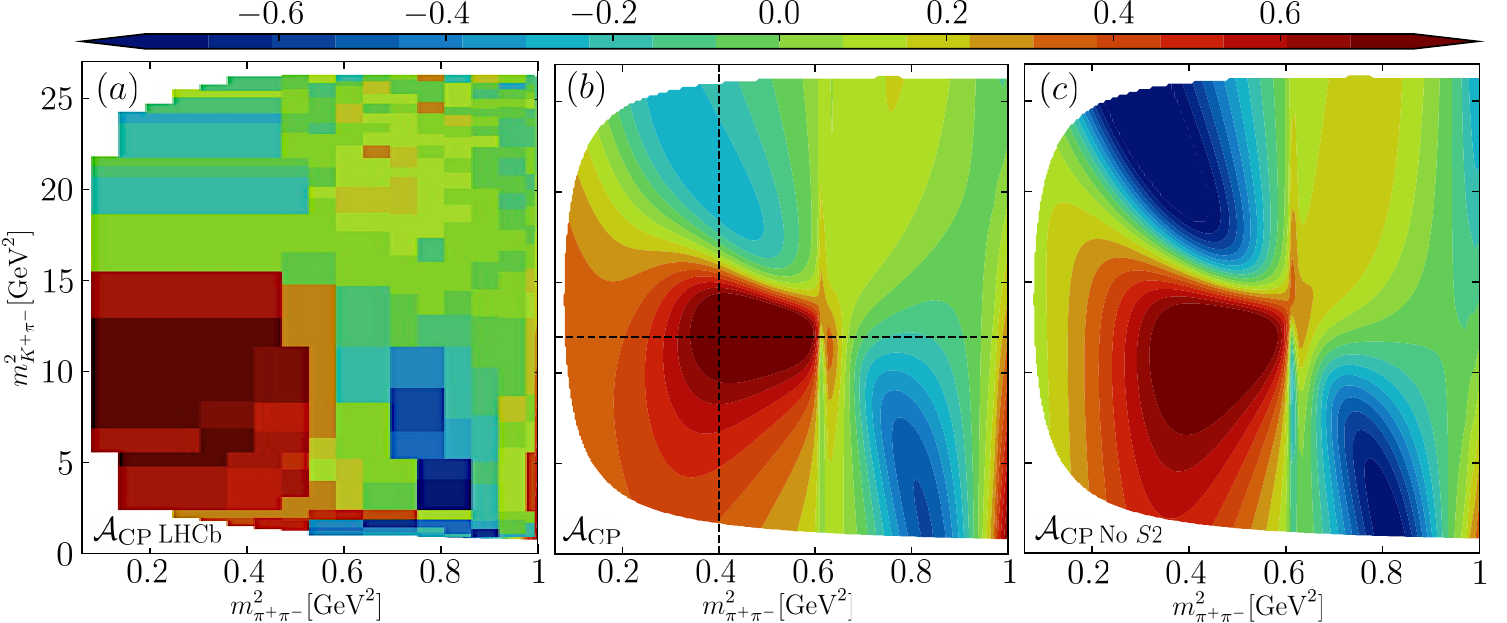}

\caption{CPV asymmetry in $B^\pm \to K^\pm \pi^+\pi^-$ decays in the
    Dalitz plot region with $s \equiv m^2_{\pi^+\pi^-} \leq 1\GeV^2$. 
    Here, $t \equiv m^2_{K^+\pi^-}$. 
    (a) LHCb raw-asymmetry results extracted from the upper right panel of Fig.~3 in Ref.~\cite{LHCb:2022fpg}; note the adaptive binning to keep the same number of events per bin. 
    (b) Predictions from our analysis using the central values of the parameters obtained from fits to projected asymmetries. The dashed vertical line is the $t$-section described in Figs.~\ref{fig:t-section} and \ref{fig:s-section} below.
    (c) The results from a fit to projected data, setting the $S2$ wave to zero.}
\label{fig:Dalitz}
\end{figure}

\section{Dalitz plot predictions }
\label{sec:Dalitzplot}
 
Finally, let us recall that we aim to describe the giant CPV asymmetry in $B^\pm\to K^\pm \pi^+\pi^-$ observed by LHCb in localized regions of their Dalitz plot analysis. With our definitions in Eqs.~\eqref{eq:mainresult} and \eqref{eq:mainresult2}, this asymmetry 
is simply:
\begin{equation}
    \ACP(s,t)=\frac{\Delta\Gamma_{\text{CP}}(s,z(s,t))}{\Sigma\Gamma(s,z(s,t))}=\frac{\Gamma^-(s,t)-\Gamma^+(s,t)}{\Gamma^-(s,t)+\Gamma^+(s,t)}.
    \label{eq:CPA}
\end{equation}
Since $\Gamma^\pm(s,t)\geq0$, it follows that  $\vert\ACP(s,t)\vert\leq1$. Note the use of variables $s$ and $t$ rather than $s$ and $z$.
The reason is that $m^2_{\pi^+\pi^-}$ and $m^2_{K^+\pi^-}$ invariant masses used by LHCb correspond to $s$ and $t$ here.

Once our parameters have been fit to the $s$ dependence of the $t$-integrated event distributions in Fig.~\ref{fig:DeltaK},
our model can predict the $t$ dependence of the CPV asymmetry~\cite{Heuser:2025mnk}. Of course, given that our approach is limited to $s\leq1\GeV^2$, in Fig.~\ref{fig:Dalitz} we can only aim at reproducing the corresponding low-$s$ section of the $\ACP(s,t)$ Dalitz plot. 
Hence, in the left panel of Fig.~\ref{fig:Dalitz}, we reproduce the LHCb result for the raw asymmetry, a plot that we have cropped and stretched from the upper right panel of Fig.~3 in Ref.~\cite{LHCb:2022fpg}. 
Note that the experimental results were obtained using adaptive binning to keep the same number of events per bin.
The central panel is the prediction of our model. 

The agreement between the data and our model is remarkable, considering that both the data and our results correspond to central values whose uncertainties are not seen in this kind of plot.
In particular, our results reproduce the two localized regions with giant CPV, i.e.,  $\vert\ACP(s,t)\vert \geq60\%$ shown in dark red or dark blue. In addition, we reproduce all the main features, including the regions where no asymmetry is observed.

\begin{figure}[t]
\centering
\includegraphics[width=0.7\linewidth]{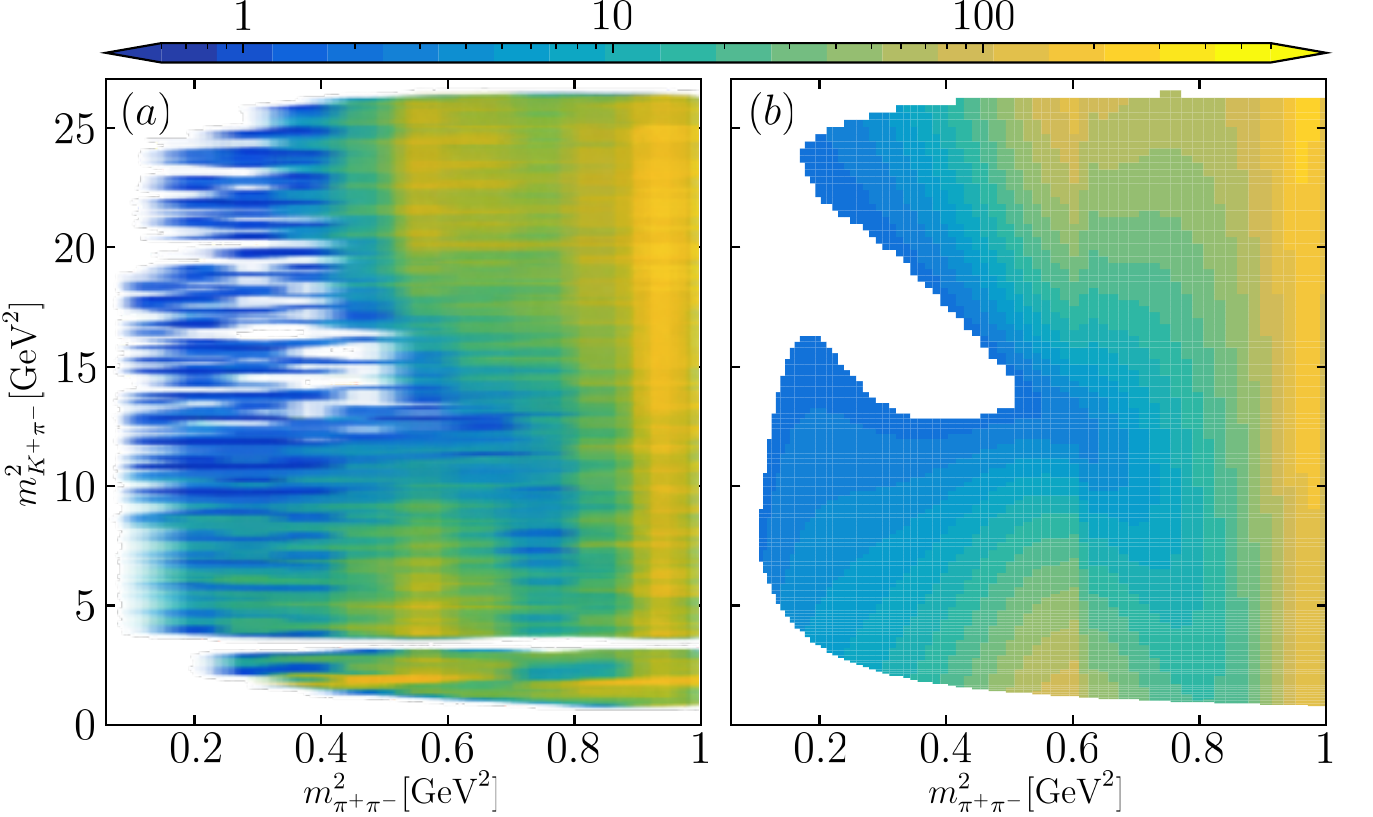}
\caption{
Left: LHCb number of events (cropped and enlarged from the upper right panel of Fig.~2 in Ref.~\cite{LHCb:2022fpg}). Right: binned predictions from our analysis using the central values of the parameters obtained from the fit to projected data. }
\label{fig:Nevents}
\end{figure}

Since $\vert\ACP(s,t)\vert$ is a ratio, one may wonder if our model can predict the absolute number of events in the $(s,t)$ plane as well. Hence, in the left panel of Fig.~\ref{fig:Nevents} we show such a number of events in the $m_{\pi^+\pi^-}<1\GeV$ region of the Dalitz plot provided by LHCb~\cite{LHCb:2022fpg}. 
Once again, this has been cropped and stretched from the upper right panel of Fig.~2 in Ref.~\cite{LHCb:2022fpg}.
In the right panel, we show our prediction using the central values of 
our model. The agreement is, once again, remarkable.

\begin{figure}
\centering
\includegraphics[width=0.7\linewidth]{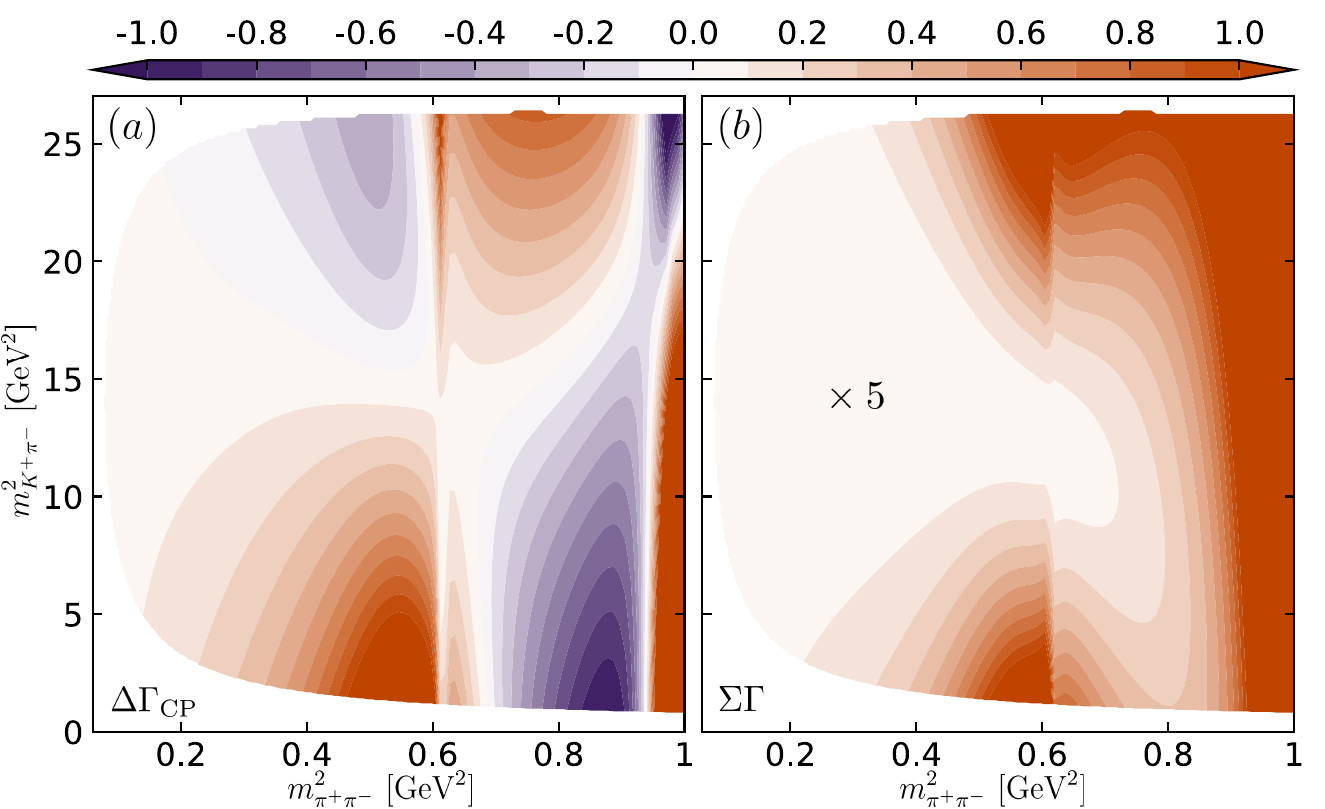}

\caption{
Numerator $\Delta\Gamma_{\text{CP}}$ (a) and denominator $\Sigma\Gamma$ (b) of our result for $\ACP$, in arbitrary units, shown in
the center panel of Fig.~\ref{fig:Dalitz}. 
Figure (b) here is the same as the right panel of Fig.~\ref{fig:Nevents} but unbinned and in a linear instead of logarithmic scale, to facilitate the comparison with (a).
}
\label{fig:NumSum}
\end{figure}

After these checks, we present the event number distribution for the numerator and denominator of the asymmetry in Fig.~\ref{fig:NumSum}. The individual $\Gamma^-$ and $\Gamma^+$ differential decay widths can be found in Fig.~\ref{fig:Amps} in Appendix~\ref{app:DPdetails}.

\section{Explaining large localized CP violations}\label{sec:largelocalCP}

\begin{figure}[t]
\centering
\includegraphics[width=10cm]{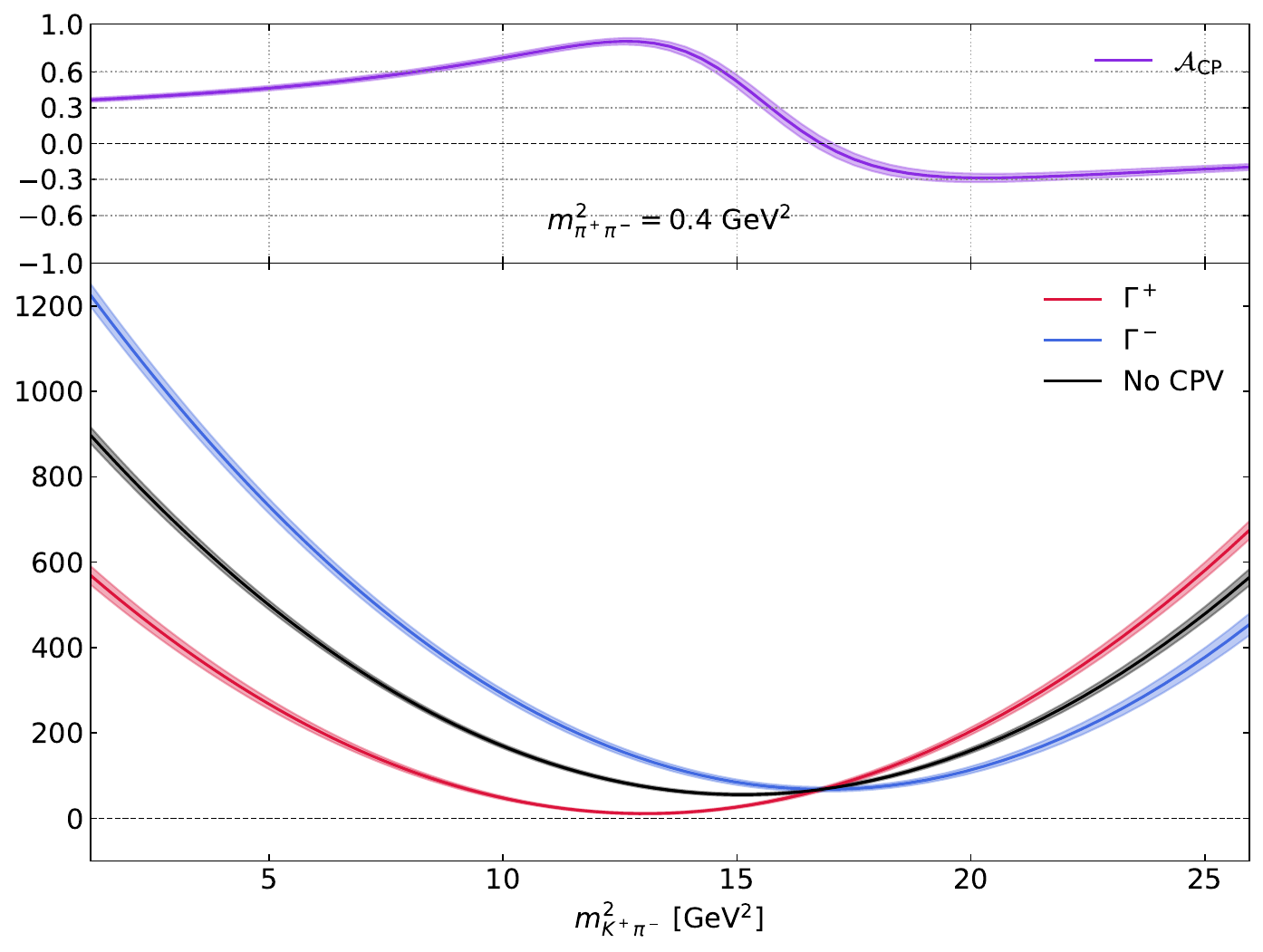}
\caption{Top: $t=m_{K^+\pi^-}^2$ dependence of the CPV asymmetry $\ACP(s_0,t)$ at fixed $m_{\pi^+\pi^-}^2 = s_0 = 0.4\GeV^2$ within our model. This corresponds to the vertical dashed line in Fig.~\ref{fig:Dalitz}. Bottom: $t$ dependence of each differential decay $\Gamma^{+}(s_0,t)$ (red) and $\Gamma^{-}(s_0,t)$ (blue). We also show their value when setting CP-violation effects to zero (black).
}
\label{fig:t-section}
\end{figure}

Once we have shown that our model predicts the asymmetries in the Dalitz plot within the elastic $\pi^+\pi^-$ regime correctly,
it is easy to understand how the large localized CPV arises. For illustration, in Fig.~\ref{fig:t-section} we first show a $t$-slice of the Dalitz plot for a fixed value of $s_0=0.4\GeV^2$ that crosses the region of largest CPV asymmetry (dashed vertical line in Fig.~\ref{fig:Dalitz}.b).
The upper panel in Fig.~\ref{fig:t-section} shows the $t=m_{K^+\pi^-}^2$ dependence of the asymmetry along this slice. Note how its value
starts at $\sim0.3$ and rises to values much larger than 0.6 in the $t=10$ to $15\GeV^2$ region, when it crosses the dark red area in the Dalitz plot. 
In order to have a large localized $\vert\ACP(s,t)\vert$, one of the $\Gamma^\pm(s,t)$
has to be much larger than the other; cf.\ Eq.~\eqref{eq:CPA}.
Thus, in the lower panel, we show both
$\Gamma^+(s_0,t)$ (red) and  $\Gamma^-(s_0,t)$ (blue), which are the result of slicing the Dalitz plots distributions in Fig.~\ref{fig:Amps} in Appendix~\ref{app:DPdetails}, as well as the common value they would have if there were no CPV (black).
The latter is nothing but half of the denominator in 
$\ACP(s,t)$, i.e., $\Sigma\Gamma(s,t)/2$.
This way, we can see the interplay between the CP-conserving contribution and 
the induced CPV difference to $\Gamma^{\pm}(s_0,t)$.
The existence of CPV is what makes the red and blue curves different and separated from the black one.
The size of both CP-conserving and -violating effects varies with $t$, i.e., with $z=\cos\theta$,
but in a different manner. In particular, the CP-conserving part reaches a minimum around $15\GeV^2$, whereas the
CPV difference vanishes, and then changes sign around $17\GeV^2$.
As a consequence, for most $t$ values the effect of CPV is smaller than the CP-conserving contribution, except around $t=8$ to $15\GeV^2$, where the black line is almost at its minimum, but the CPV is not. This is where we find a large, $>0.6$, localized asymmetry.
Therefore, it is not that CPV effects in this region become larger in absolute size than in other regions, but that the CP-conserving part is suppressed here, to the point of becoming similar in size to the CPV contribution. We have checked with other slices that this is the generic pattern when a large localized CP violation is observed.

\begin{figure}[t]
\centering
\includegraphics[width=10cm]{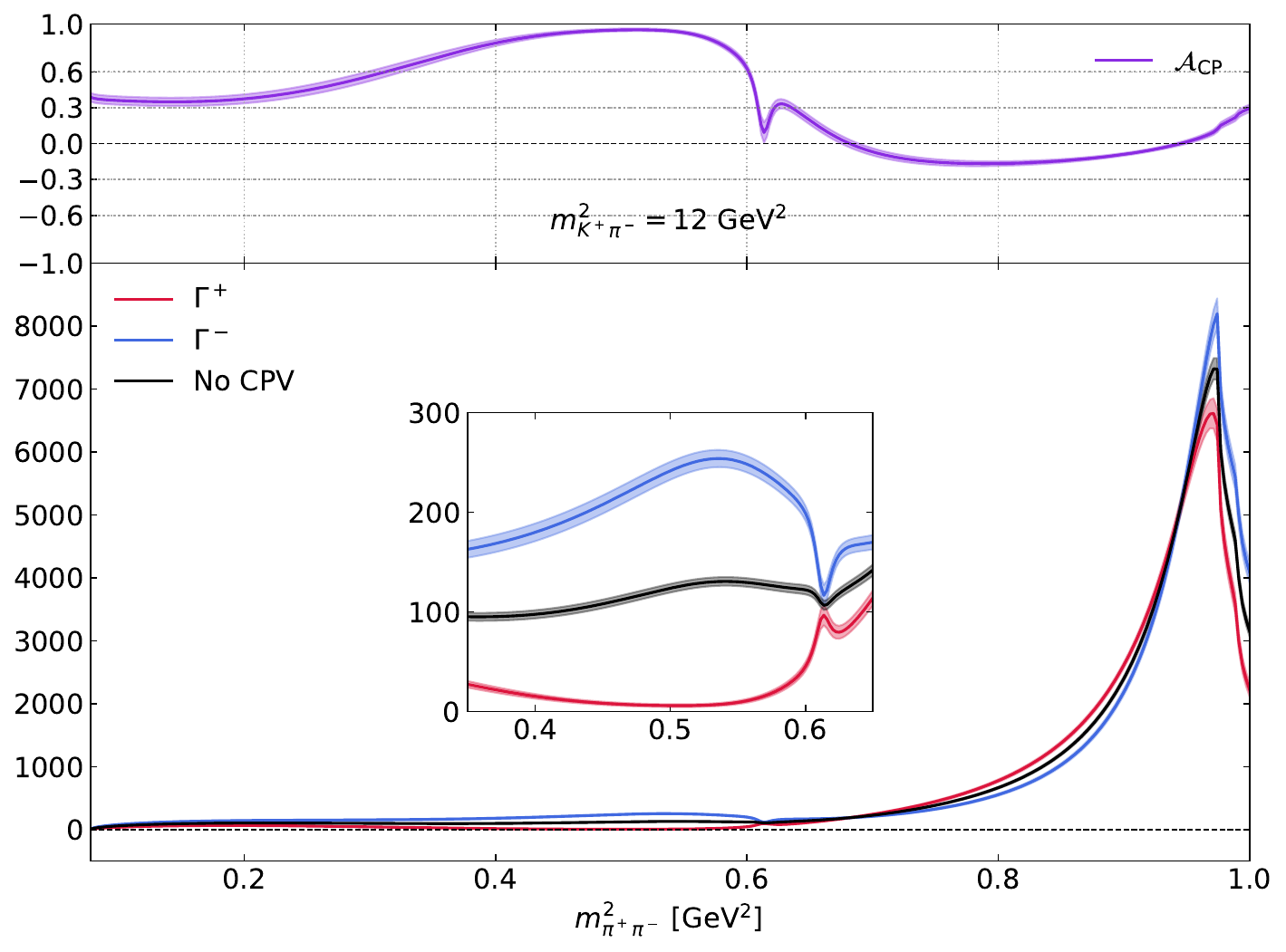}
\caption{ Top: $s=m_{\pi^+\pi^-}^2$ dependence of the CPV asymmetry $\ACP(s,t_0)$ at fixed $m_{K^+\pi^-} = t_0 = 0.12\GeV^2$ within our model. This corresponds to the horizontal dashed line in Fig.~\ref{fig:Dalitz}. Bottom: $s$ dependence of each differential decay $\Gamma^{+}(s,t_0)$ (red) and $\Gamma^{-}(s,t_0)$ (blue). We also show their value when setting CP-violation effects to zero (black).
}
\label{fig:s-section}
\end{figure}

In addition, in Fig.~\ref{fig:s-section}, we show an $s$-slice of the Dalitz plot for a fixed value of $t_0=0.12\GeV^2$ that also crosses the region of largest CPV asymmetry (dashed horizontal line in Fig.~\ref{fig:Dalitz}).
The upper panel describes the $s=m_{\pi^+\pi^-}^2$ dependence of the asymmetry along that slice, which reaches a value close to 1 in between $s=0.4$ and 0.6~GeV$^2$. In the lower panel, we now see the individual $\Gamma^\pm(s,t_0)$ contributions and their common value when CPV effects are turned off. The most remarkable feature is the very large peak right below $s=1\GeV^2$, clearly related to the $f_0(980)$ resonance. Such a
peak is also present in the $t$-integrated $\Sigma\Gamma^{\pm}$ in Fig.~\ref{fig:DeltaK}. However, 
in that figure, we also see a peak near the $\rho(770)$ resonance, which is not seen in Fig.~\ref{fig:s-section} around $\sim(0.775\GeV)^2\sim0.6\GeV^2$. 
Even in that region's detailed inset plot, we see a large cancellation of this resonance
 due to its interference with other waves.  Consequently, the CP-conserving contribution is similar in size to the CPV contribution.  As a result, $\Gamma^+$ almost cancels, leading to a very large and localized $\ACP$. Thus, it is not the resonance peak that yields the giant $\ACP$, but the suppression of the CP-conserving part due to its interference with other partial waves.

\section{The importance of the \texorpdfstring{\boldmath{$S2$}}{S2} wave}
\label{sec:S2effect}

Let us now discuss the effect of removing the $S2$ wave in detail, which is usually omitted in the literature. A possible reason for that omission is that decays are dominated by resonance contributions,  but this is a non-resonant wave. However, in the study of CPV, we are not so much interested in total yields or total number of events, but in the difference between CP conjugated processes. These are often driven by the different phases between two waves.
 Indeed, we already saw in Fig.~\ref{fig:ContributionsImportant} that interferences with the $S2$ wave provide relevant contributions to the integrated CPV differences.

\begin{figure}[t]
\centering
\includegraphics[width=\textwidth]{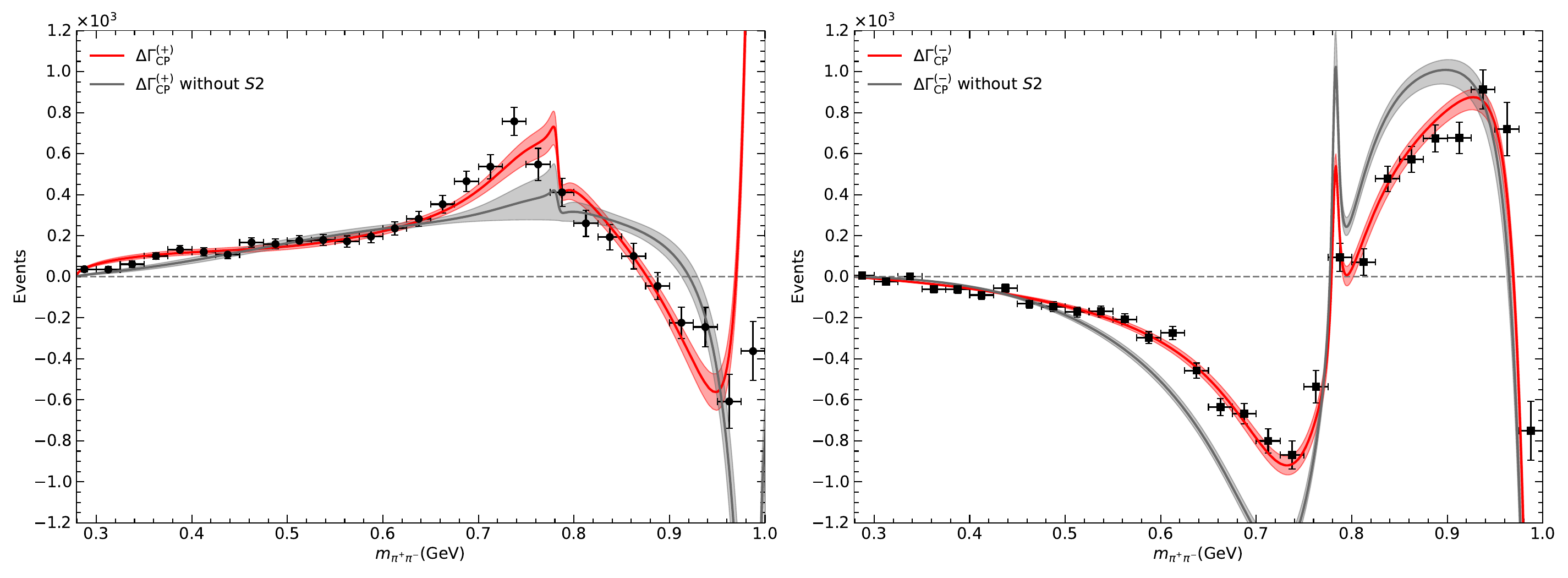}
\caption{ Effect of fitting the projected data without $S2$-wave contribution. We show the resulting
CP differences $\Delta\Gamma_{\text{CP}}^{(\pm)}(s)$.  
By comparing with the full model in the lower panels of Fig.~\ref{fig:DeltaK}, we see that, ignoring the $S2$ wave, the data description becomes considerably worse.
}
\label{fig:NoS2}
\end{figure}

Actually, in Fig.~\ref{fig:NoS2} we show the fit
to the projected CP differences defined in Eqs.~\eqref{eq:diffFBsym} and \eqref{eq:diffFBasym} 
when omitting the $S2$ wave in gray. It is clear that the data fit becomes much worse than that of the full model, which we show again for comparison (in red). By ignoring the $S2$ wave, the fit fails to reproduce salient features like the peak in $\Delta\Gamma^{(+)}_{\text{CP}}$, while overestimating the CP violation in  $\Delta\Gamma^{(-)}_{\text{CP}}$.

Concerning ${\cal A}_{\text{CP}}(s,t)$, in the right panel of Fig.~\ref{fig:Dalitz} we show the distortion in the Dalitz plot prediction when the $S2$ wave is omitted in the fit of the integrated data. In particular, the negative asymmetry in the upper left region of the Dalitz plot becomes much more pronounced than in the data (left panel).

\section{Comparison with other formalisms}
\label{sec:literature}

There are alternative implementations of the separation of three-body decay amplitudes into FSI and source terms in the literature.

A particularly popular approach to FSI in $B$-meson decays is based on the Suzuki--Wolfenstein (SW) formalism~\cite{Suzuki:1999uc,Chua:2007cm}, which we discuss in detail in Appendix~\ref{app:otherapproaches}.
Our method has several advantages over the SW approach. In brief, in the SW framework, the Omnès function or matrix (along with the polynomial) in Eqs.~\eqref{eq:Aelast} and~\eqref{eq:2channelMOs} is replaced by the $\mathcal{S}^{1/2}$ matrix. We highlight the better analytic properties of our approach and the clean factorization of FSI, which is often compromised in the SW method. Moreover, because evaluating this matrix square root is inherently difficult without resorting to unrealistic approximations~\cite{Cheng:2016shb,Cheng:2020ipp}, the SW formalism is typically restricted to leading order within various inelasticity expansions~\cite{Wolfenstein:1990ks, Bediaga:2013ela, AlvarengaNogueira:2015wpj,Garrote:2022uub} (see Ref.~\cite{Bigi:2000yz} for a textbook introduction). Truncating the $\mathcal{S}$-matrix expansion in this manner means that Watson's theorem is satisfied only approximately, whereas such approximations are entirely unnecessary in our framework.

Concerning the sources, in the exact $m_b\to\infty$ limit, our source terms can be evaluated within the quasi-two-body factorization framework.
This framework attempts to generalize the two-body factorization ansatz, where the matrix element of four-quark operators between the heavy-meson initial state and the two-light-meson final state factors into 
the heavy-to-light transition form factor from the heavy meson to one light meson 
and the convolution of the hard scattering kernel and the light-cone distribution amplitude of another light meson, often written as the decay constant times the factorization coefficient.
It is common in the literature~\cite{Furman:2005xp, El-Bennich:2006rcn, El-Bennich:2009gqk} to omit the dependence of the two-meson light-cone distribution amplitude on the invariant mass other than the local form factor, which serves as normalization.
While the initial factorization ansatz was conjectured based on a next-to-leading-order evaluation of partonic matrix elements~\cite{Beneke:1999br}, progress to associate the ansatz with a specific set of diagrams had been made for the two-body case by Chay and Kim~\cite{Chay:2003ju}, and for the three-body case 
in the restricted kinematic regions of the Dalitz plot where a collinear two-meson pair recoils against the bachelor particle,
by Kränkl, Mannel, and Virto~\cite{Krankl:2015fha}, where the role of light-cone distribution amplitudes is kept explicit.  

Straightforward implementations of the quasi-two-body factorization ansatz, however, result in branching ratios 2--5 times below the observation~\cite{Cheng:2007si,El-Bennich:2009gqk}.
El-Bennich \emph{et al.}~\cite{El-Bennich:2006rcn,El-Bennich:2009gqk} therefore add phenomenological penguin parameters to the hard scattering kernel, while still omitting the $S2$ wave as well as higher partial waves of the collinear two-meson pair. Note that in Ref.~\cite{El-Bennich:2009gqk}, the authors use a  Muskhelishvili--\Omnes formalism for their $K\pi$ form factors, but the non-resonant $I=3/2$ partial waves were ignored.
Boito \emph{et al.}~\cite{Boito:2017jav} reformulated the structure with a more phenomenological stance, absorbing the factorization coefficients and penguin parameters together into complex coefficients. 
As one alternative to describe the $S0$ form factor, they  use the $\pi\pi$--$K\bar K$ coupled-channel Muskkelishvili--\Omnes formalism, following Ref.~\cite{Moussallam:1999aq}.

In our dispersive construction, the treatment of the source terms departs from this quasi-two-body factorization philosophy in several important respects:
\begin{enumerate}
\item We do not attempt to keep track of the operator basis at the amplitude level. Instead, we collect the entire set of operators of the 
effective weak Hamiltonian, with all associated Wilson coefficients, into a handful of effective sources for each partial wave.  In this language, the weak phase $\gamma$ appears only through the Standard Model $\bar{b} W u$ vertex, while the remaining short-distance information from tree, QCD penguin, and electroweak penguin topologies is absorbed into real coefficients.
Furthermore, we do not refer to the factorization ansatz for the evaluation of QCD amplitudes.
\item Second, most of the energy dependence and in particular the strong phases are generated by universal two-body rescatterings, described by Omnès functions and Omnès matrices, constrained by high-accuracy information on $\pi\pi$ scattering. This ensures that strong phases are not model parameters but fixed inputs, making the resulting CP-asymmetry distributions genuine predictions of the formalism once the source strengths are determined.

Moreover, our formalism also leaves room for an additional regular energy dependence allowed by two-body unitarity, which we parametrize with a polynomial. While the simplified factorization ansatz fixes energy dependencies as local form factors, our parametrization allows for a consistency check that the energy dependence is indeed mostly captured by the \Omnes matrix. 
\item Third, we attribute the CP-even absorptive parts, crucial for describing the large localized CP asymmetries observed experimentally, naturally to an additional charm-rescattering layer beyond the weak vertex. 
While the physics behind the phenomenological penguin parameters,
introduced to supplement the perturbatively evaluated charm-loop effects in the evaluation of the QCD hard-scattering kernel within the quasi-two-body factorization ansatz,
is left obscure~\cite{El-Bennich:2006rcn, El-Bennich:2009gqk}, 
in our setup, the effective weak vertex
is followed by hadronic rescattering.
The CP-even imaginary parts of the source term
should thus be
interpreted as intermediate $D^{(*)}\bar D_s^{(*)}$ states that eventually
connect back into the light three-body system---for  typical diagrams see Fig.~\ref{fig:triangle}.
 The effect of the same dynamics, but in a different area of the Dalitz plot, was studied in
Ref.~\cite{Bediaga:2020ztp}.
\item Finally, we allow isospin-2 source terms, followed by non-resonant final-state interaction. Because of the absence of resonances in the $\pi\pi$ isospin-2 final state, and since in QCD factorization 
such terms are formally power suppressed~\cite{Krankl:2015fha}, they were neglected in earlier studies of $B$ decays into light hadrons. However, they proved crucial for a successful description of the data.  
\end{enumerate}

\section{Summary and outlook}
\label{sec:summary}

In this paper, we provided an in-depth study of a novel dispersive framework for the analysis of CP asymmetries in hadronic multi-body $B$ decays, proposed in Ref.~\cite{Heuser:2025mnk}. For illustration, the formalism is applied to the $B^\pm \to K^\pm \pi^+ \pi^-$ channel in the kinematic regime
where the two-pion pair is elastic ($m_{\pi\pi} < 1 \GeV$). By employing a model-independent approach based on unitarity and analyticity, we successfully addressed the limitations of traditional isobar models and quasi-two-body factorization schemes.

The main findings and phenomenological implications of this work are as follows.

\begin{itemize}

    \item Our formalism establishes a transparent connection between 
    the short-range weak
    transition and the long-range rescattering dynamics, captured model-independently by the dispersively reconstructed partial waves. By parametrizing the source terms as linear polynomials and accounting for {charmed-meson loops} via constant CP-even imaginary parts, we reproduce the rich structure of the Dalitz plot with remarkable agreement, using only a handful of physically motivated parameters.

    \item The inclusion of the {isotensor $S2$~wave} is shown to be essential for a successful description of the data. Although non-resonant, the $S2$ wave provides critical interference terms that significantly improve the fit quality of the forward--backward symmetric and antisymmetric CP differences, $\Delta \Gamma_{\text{CP}}^{(\pm)}$.
    
    \item We demonstrated that the {large localized CP asymmetries} ($\vert\ACP\vert \geq 60\%$) reported by LHCb typically do not arise from an absolute enhancement of CP-violating effects. Instead, our analysis reveals that they are primarily driven by the locally small CP-conserving parts of the amplitude, resulting from the intricate interference between different partial-waves in specific regions of the Dalitz plot.
    
\end{itemize}

The success of the method in describing the low-mass region of $B^\pm \to K^\pm \pi^+ \pi^-$ provides a clear roadmap for future extensions. The modularity of the dispersive construction allows for the inclusion of inelastic thresholds, such as $\pi\pi \to K\bar{K}$ at higher energies, and its adaptation to other decay modes like $B^\pm \to 3\pi$ or $b$-baryon decays. Given the increasing  precision
of the data sets from LHCb and Belle II, the systematic treatment of hadronic interactions presented here will be vital for performing high-precision tests of the Standard Model and searching for or discarding new sources of CP violation.

\acknowledgments
This work was initiated and formulated during the three-month research visit of J.R.P.\ to Bonn University under the Spanish MCIN ``Salvador de Madariaga'' Program, Grant No.~PRX22/00129, as well as during the preparations of the Color meets Flavor cluster, funded by the Deutsche Forschungsgemeinschaft (DFG, German Research Foundation) under Germany's Excellence Strategy -- EXC 3107 -- Project-ID 533766364. Financial support by the MKW NRW under funding code NW21-024-A, by the Konrad-Adenauer-Stiftung e.V.\ with funds from the BMBF, the Spanish Grant PID2022-136510NB-C31 funded by MCIN/AEI/10.13039/501100011033, the European Union's Horizon 2020 research and innovation program under grant agreement No.~824093 (STRONG2020), the Brazilian funding FAEPEX grant 2030-24, and INCT-FNA CnPq grant, is gratefully acknowledged. T.M.\ was supported by the Deutsche Forschungsgemeinschaft (DFG, German Research Foundation) under Grant No.~396021762-TRR 257 ``Particle Physics Phenomenology after the Higgs Discovery.'' C.H. thanks the CAS President's International Fellowship Initiative (PIFI) under Grant No.\ 2025PD0087 for partial support.

\appendix

\section{Fit details}\label{app:fitparams}

In this appendix, we detail our fitting procedure and the resulting best-fit parameters. The data and associated uncertainties were extracted directly from the figures in Ref.~\cite{LHCb:2022fpg} (supplemental material).

\subsection{Strategy}

In this work, we focus on the elastic region for the outgoing pion pair. Accordingly, we fit the LHCb data for the projected $B^+$ and $B^-$  yields only up to an invariant mass of $E \equiv \sqrt{s} = E_\text{max}=1\GeV$. To take the finite bin size of the experiment into account, we  use 
 \begin{equation}
     \tilde{f}(E_i,z)=\frac{1}{2\delta E}\int_{E_i-\delta E}^{E_i+\delta E}\frac{\diff^2\Gamma^\pm}{\diff\sqrt{s}\,\diff z}(E^2,z) \,\text{d}E,
 \end{equation}
 with $2\delta E=25\,\text{MeV}$ for the width of the bins.
 Since the data for the projected yields are neither 
 acceptance corrected nor background subtracted, we apply
 the method of Ref.~\cite{LHCb:2022fpg} to correct
 for the former and allow for a linearly rising, non-interfering
 background to account for the latter. 
 Thus we minimize
 \begin{equation}
     \chi^2=\sum_i \left(\tilde{f}(E_i,z)+ B(\sqrt s-2M_{\pi}) -N_i\right)^2/\sigma_i^2 \,,
 \end{equation}
where the slope parameter is set to $B=\SI{50\pm20}{GeV^{-1}}$, intended to mimic the combinatorial background projections utilized by LHCb in their ongoing analysis~\cite{LHCb:privcomm} and Ref.~\cite{LHCb:2022fpg}. The uncertainty of the background is absorbed into that of each respective bin.

Our best fit yields a $\chi^2/\text{d.o.f.}=3.4$. However, beyond potential concerns regarding the precision of extracting data from figures, we contend that this $\chi^2/\text{d.o.f.}$ lacks a strict statistical interpretation. This is because the projected data have not been corrected for acceptance effects (such as efficiency and production asymmetry) nor have they undergone background subtraction, although, as we have just explained, we have done our best to mimic such effects. 

It is worth noting that the LHCb collaboration does not typically provide $\chi^2/\text{d.o.f.}$ values for their Dalitz-plot amplitude analyses or their projections (see, for instance, the $B\to3\pi$ analysis in Ref.~\cite{Aaij:2019jaq}). Similarly, previous theoretical studies of Belle~\cite{Belle:2005rpz} and BaBar~\cite{BaBar:2008lpx} data—such as those in Refs.~\cite{Simma:1991ct,Fajfer:2004cx,Furman:2005xp,El-Bennich:2006rcn,El-Bennich:2009gqk}—fit the projections directly without reporting corresponding $\chi^2/\text{d.o.f.}$ values.

\subsection{Parameters and covariance}

\begin{table}[t!]
\centering
\begin{tabular}{c l}
\toprule
Parameter & Value \\
\midrule
$a_{S0n}$ & $\phantom{-}\SI{6.8(1.3)e2}{} $ \\

$a_{S0s}$ & $\SI{-8.9(7)e3}{}$ \\

$a_{S2}$ & $\SI{-2.6(3)e3}{}$ \\

$a_{P}$ & $\SI{-5.9(3)e1}{GeV^{-2}}$ \\

$b_{S0n}$ & $\phantom{-}\SI{1.0(1)e3}{}$ \\

$b_{S2}$ & $\phantom{-}\SI{2.9(1)e3}{}$ \\

$b_{P}$ & $\SI{-5.0(3)e1}{GeV^{-2}}$ \\

$c_{S0n}$ & $\SI{-6.0(1)e2}{}$ \\

$c_{S0s}$ & $\SI{-2.4(3)e3}{}$ \\

$c_{P}$ & $\phantom{-}\SI{0.53(8)e1}{GeV^{-2}}$ \\

$p_{S0}$ & $\SI{-0.41(6)}{GeV^{-2}}$ \\

$p_{S2}$ & $\SI{-1.67(3)}{GeV^{-2}}$ \\

$p_{P}$ & $\phantom{-}\SI{0.4(1)}{GeV^{-2}}$ \\
\bottomrule
\end{tabular}
\caption{Parameter of the fit. The absolute values absorb a normalization obtained by fitting the number of events provided by LHCb in their angle-integrated data. It is the relative sizes and signs that matter.
}
\label{tab:tab:fit-params_updated}
\end{table}

\begin{figure}[t!]
    \centering
    \includegraphics[width=0.9\linewidth,trim=5cm 9cm 7cm 7cm ,clip]{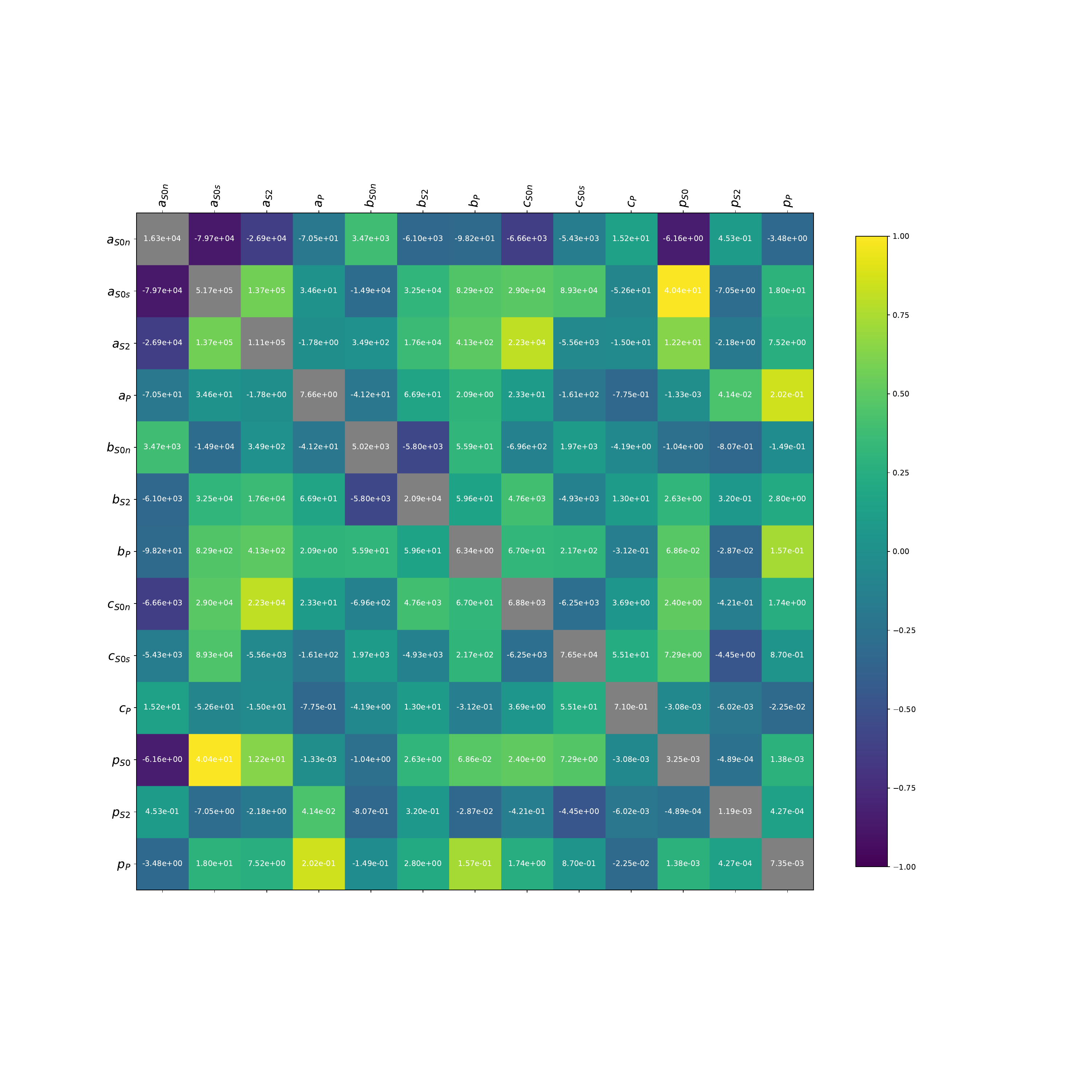}
    
    \caption{Covariance matrix of the results. Color indicates the correlation.}
    \label{fig:covariance_matrix}
\end{figure}

We are fitting to yields given as $\text{yields}/\SI{0.025}{GeV}$, which imposes a normalization factor $N=\SI{0.025}{GeV}$ on our fit parameters. Taking this into account, our parameters can be rescaled following the prescription
\begin{equation}
    \hat{x_i}=x_i\cdot\sqrt{1/N},
\end{equation}
with $x\in\{a,b,c\}$ and $i\in\{S0n,\, S0s,\, S2,\, P\}$ and $\hat{x}$ being the proper parameter, which is connected directly to the weak effective operators. 
Furthermore, we provide the covariance matrix of our best fit parameters in Fig.~\ref{fig:covariance_matrix}.

\subsection{Stability of the result}
In the approach presented above, we omit some details about the Dalitz distribution that can potentially influence the fit in a non-negligible way. To understand the effect these potential sources of uncertainty can have on the fit, we carry out extensive stability studies on our results.

\begin{figure}[t]
    \centering
    \includegraphics[width=\textwidth]{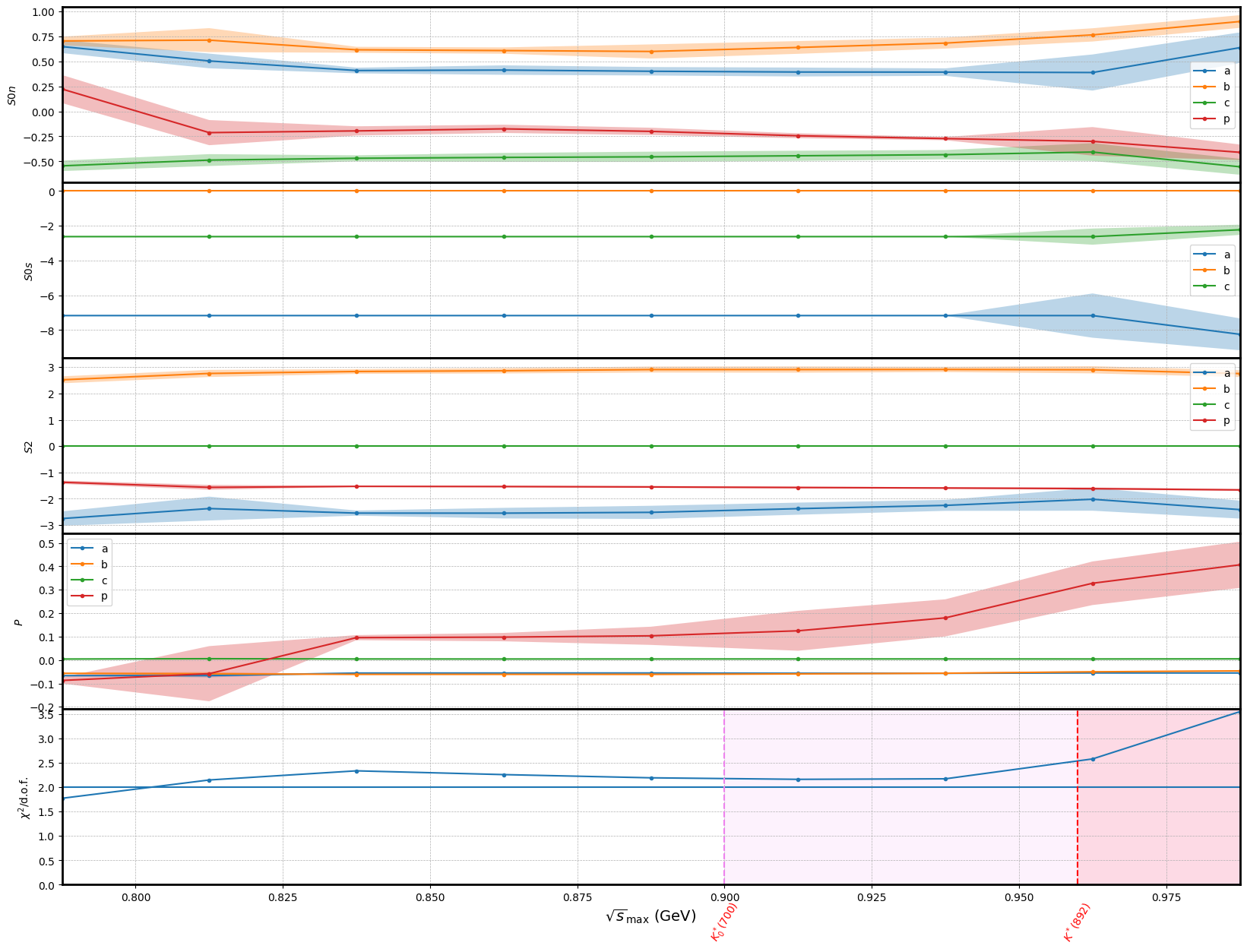}
    \caption{Results of the stability test in which the fit range is varied downwards. Each partial wave has its own diagram with all the parameter changes shown. Color bands indicate the error Minuit provides for that fit. Also indicated are the minimal values of $\sqrt{s}=m_{\pi\pi}$ corresponding to crossed-channel $K\pi$ resonances (using the central value of $\mathrm{Re}(\sqrt{u_P})$).}
    \label{fig:range_stability}
\end{figure}

\begin{figure}[t]
    \centering
    \includegraphics[width=\textwidth]{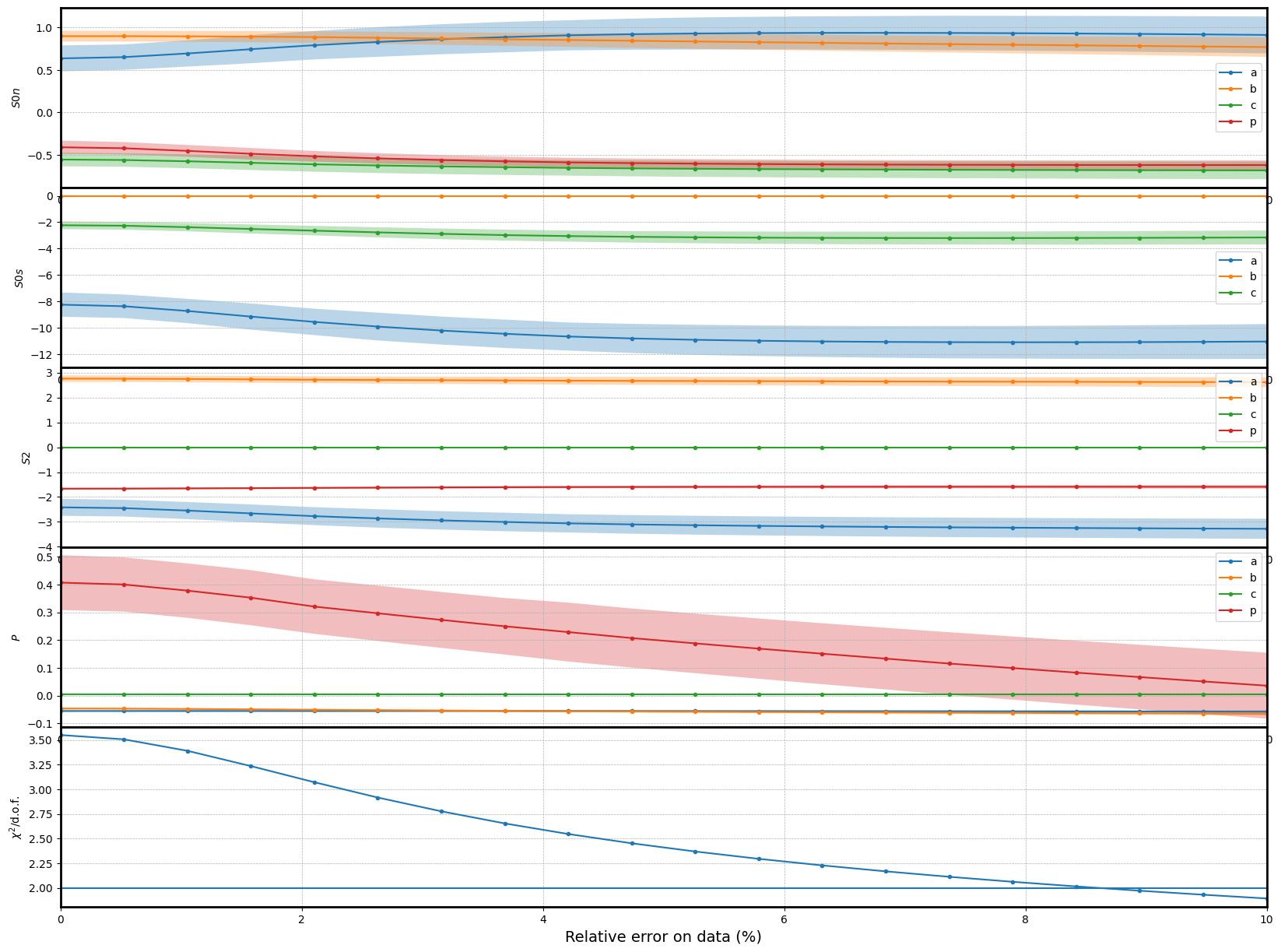}
    \caption{Result of the stability test in which an increasingly large relative error is added to the data. Each partial wave has its own diagram with all the parameter changes shown. Color bands indicate the error Minuit provides for that fit.}
    \label{fig:rel_error_stability}
\end{figure}

The main sources of systematic uncertainty are crossed-channel $K\pi$ resonances and the $D^0$ cut ($1.740\,\text{GeV}<m_{K\pi}<1.894\,\text{GeV}$)~\cite{Aaij:2019qps} that was applied to the data, since we omit these effects completely. While to take into account the crossed-channel resonances would necessitate a different framework, the $D^0$ cut is treatable within our current ansatz. To implement the effect of the cut, we have to correct the angular integration in Eq.~\eqref{eq:angular_integration} accordingly. The two limits of the cut in $m_{K\pi}^2=u$ translate into $m_{\pi\pi}^2=s$ dependent cosine limits 
\begin{equation}
    \cos{\theta}_j(s)\approx\frac{u_j-\frac{M_B^2}{2}}{g(s)},
\end{equation}
with $j=1,\,2$ being the two limits.  We have to subtract the angular integration in these regions,

\begin{equation}
\delta_{ij}=\int_{\cos{\theta}_1(s)}^{\cos{\theta}_2(s)}f_i(s,z)f_j(s,z)\text{d} z  \, ,  
\end{equation}
from the full integral to account for the missing data. This affects the orthogonality relation of the different partial waves and causes additional contribution of partial-wave interferences in the different $\Delta\Gamma^{(\pm)}$ and $\Sigma\Gamma^{(\pm)}$. Taking into account these effects, we only observe a minimal change in the $\chi^2$.

Further tests of stability are performed as follows. The first test is carried out by varying the range in which we apply our fit. The range is decreased systematically and the newly fitted parameters are used as the start point for the next fit. Below twice the width of the $f_0(980)$, the $S0s$ parameters are frozen to avoid artifacts, since the $f_0(980)$ contribution cannot be properly fixed anymore. From this strategy one is potentially able to infer information about the influence of different $K^*$ resonances in $m_{K\pi}^2$, since depending on their masses their influence is limited to different regions of $m_{\pi\pi}^2$.

Furthermore, a second strategy is applied in which a rising relative error is added to the data. Again the refit parameters will be the starting point of the next step. This test will enable us to judge the general stability of the fit parameters. This might also be relevant in the context of the missing part of the Dalitz plot that was cut because of the $D^0$. The results of both studies can be seen in Figs.~\ref{fig:range_stability} and \ref{fig:rel_error_stability}. As can be inferred from the given plots, all parameters stay stable within at most $1.5\sigma$ of their error band. Interestingly, we see clear improvements in the $\chi^2$ while decreasing the fit range. This is a possible hint towards the influence of crossed-channel resonances that play a less important role the further one decreases the range of $s$. However, it could also be just a statistical effect from the exclusion of some points that do not fit in our formalism. 

We tested the additional implementation of the isoscalar $D$ wave. However, within the data there is no clearly visible resonance signal around the mass of the $f_2(1270)$. For that reason, we imposed constraints on the parameters of the $D0$ wave to limit the strength of the signal. Fitting within those constraints does not result in any improvement of the $\chi^2/\text{d.o.f.}$. The parameters of the other waves only change by up to $1.5\sigma$.

\section{Contributions to Dalitz plots}
\label{app:DPdetails}

In this appendix we provide further details on our results for the $\Acal_{\text CP}$ Dalitz plot and its contributions. 

\begin{figure}[t]
\centering

\includegraphics[width=\textwidth]{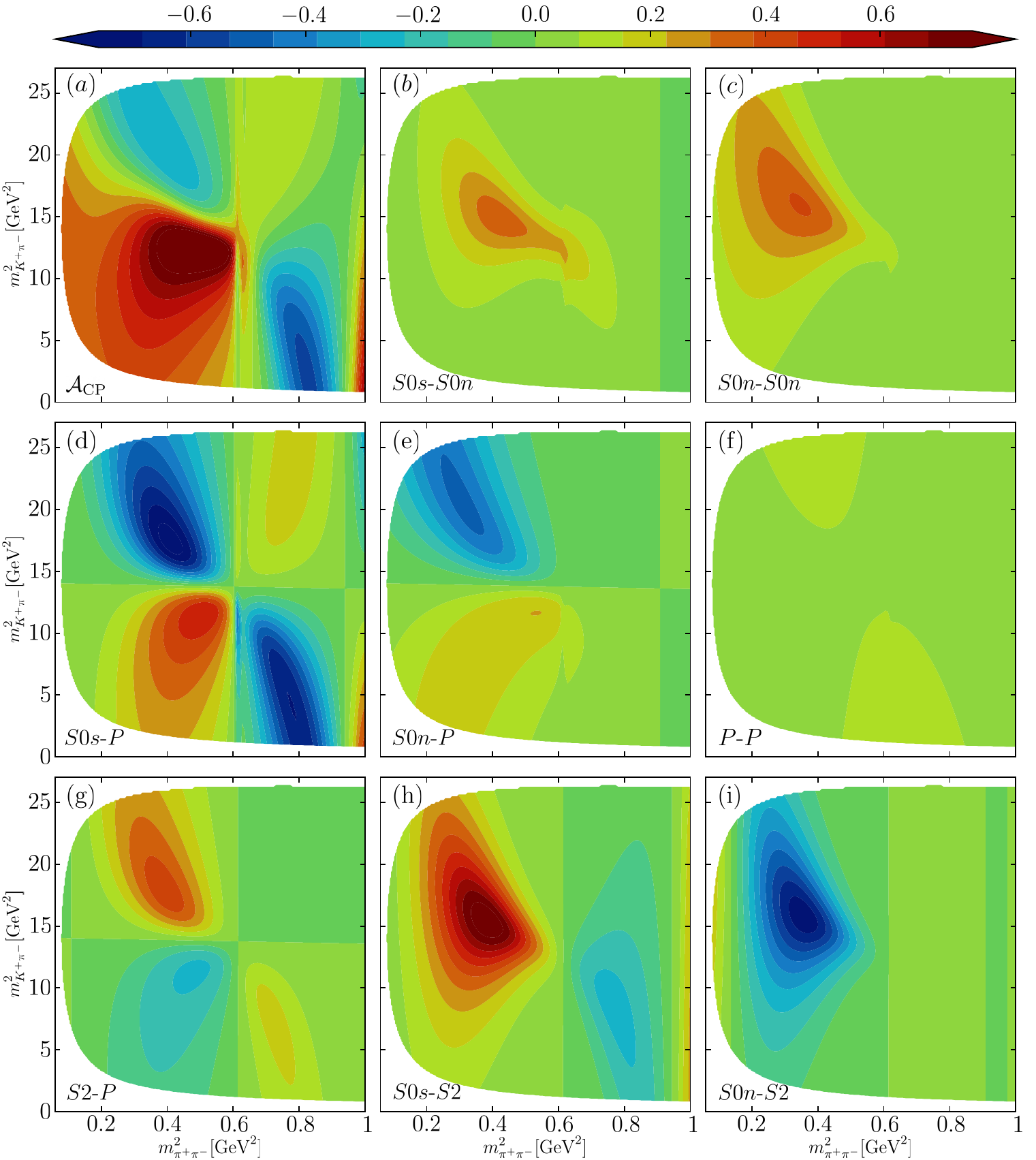}

\caption{
    Dalitz plot section, with $s \equiv m^2_{\pi^+\pi^-} \leq 1\GeV^2$, of the different interference contributions to the CPV asymmetry in $B^\pm \to K^\pm \pi^+\pi^-$ decays. (a) Complete CPV-asymmetry prediction from our analysis using the central values of the central values of the parameters. Panels (b)--(i) display each contribution separately.}
\label{fig:ContAsym}
\end{figure}

\begin{figure}[t]
\centering

\includegraphics[width=\textwidth]{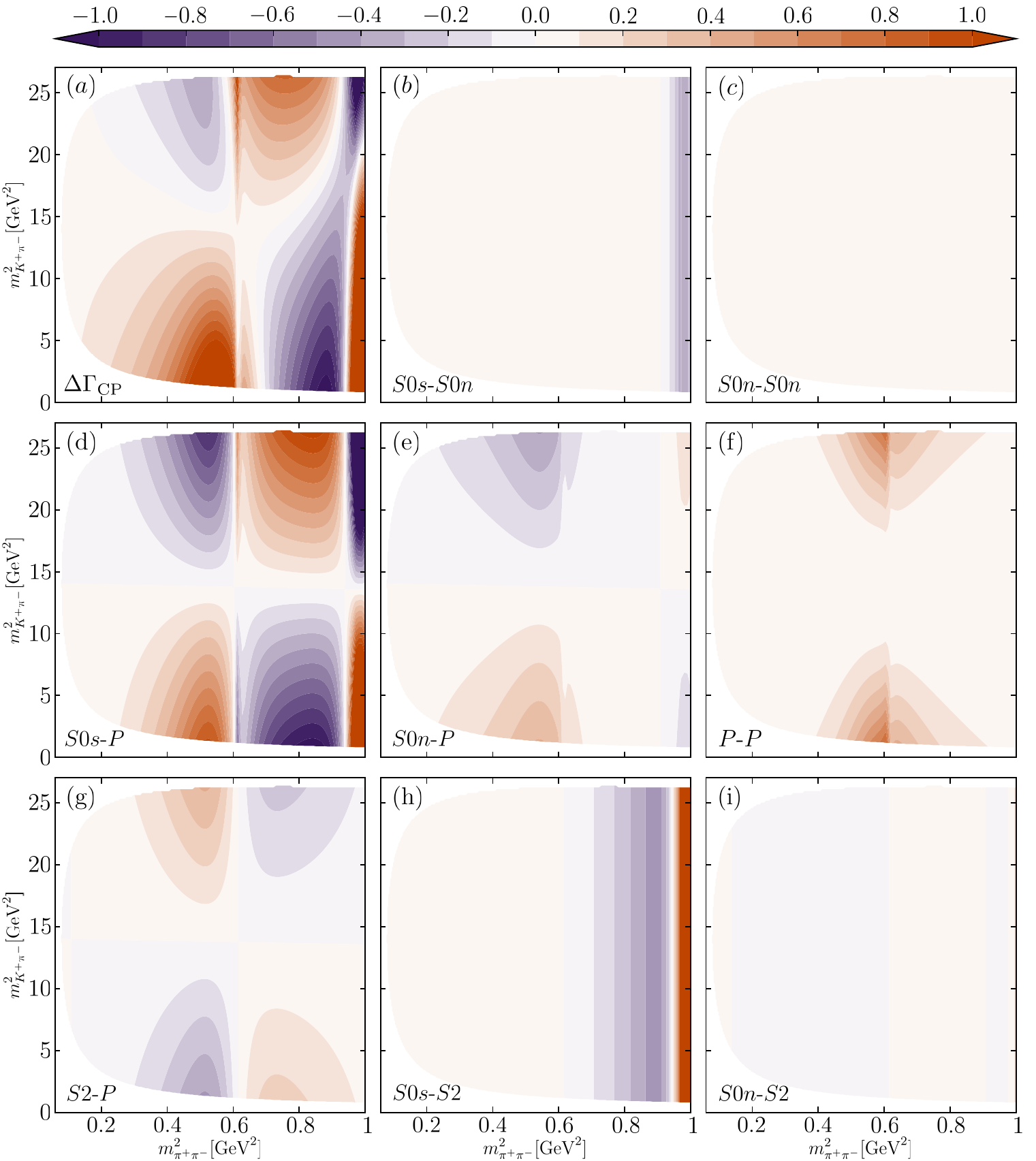}

\caption{
    Dalitz plot section, with $s \equiv m^2_{\pi^+\pi^-} \leq 1\GeV^2$, of the different interference contributions to the numerator $\Delta\Gamma_{\text{CP}}$ of the CPV asymmetry in $B^\pm \to K^\pm \pi^+\pi^-$ decays. (a) Complete numerator $\Delta\Gamma_{\text{CP}}$ of CPV-asymmetry prediction from our analysis using the central values of the parameters. Panels (b)--(i) display each contribution separately. Note the main features are provided by the (d) $S0s$--$P$ term.}
\label{fig:ContNum}
\end{figure}

In particular, Fig.~\ref{fig:ContAsym}(a) displays the full asymmetry $\Acal_{\text CP}$ predicted by this method, also shown in Fig.~\ref{fig:Dalitz}(b) of the main text. Panels (b)--(i) show its different contributions. To ease the comparison we make them share the same common denominator, $\Sigma\Gamma(s,t)$, so that their sum reproduces the full $\Acal_{\text CP}$. Each panel contains the interference or modulus square contribution of the partial waves $i$--$j$, with $i,j\in\{S0n, S0s, S2, P\}$, as defined in the main text. It can be observed that $\Acal_{\text CP}$ cannot be attributed to a single term; rather, only the combined effect of all of them reproduces the experimentally observed structure. However, our use of a common CP-symmetric denominator with the sum of all contributions, which has its own $(s,t)$ dependence, may obscure the distribution of pure CPV effects.

Hence, to disentangle pure CPV effects, in Fig.~\ref{fig:ContNum} we also provide the corresponding numerators of the total $\Acal_{\text CP}(s,t)$ (a),  and the individual contributions (b)--(i). The $S0s$--$P$ interference, shown in panel (d), captures the main features of the full numerator. The role of the $S2$ wave is also very significant, as the $S2$--$P$ contribution in panel (g) provides the necessary compensation to the $S0s$--$P$ term.

Furthermore, each $i$--$j$ panel in Fig.~\ref{fig:ContNum} can be related to the projected contributions shown in Fig.~\ref{fig:ContributionsImportant} of the main text. 
It is worth noting that although the $P$--$P$ effect in panel (f) is essential to reproduce the $\Delta\Gamma_{\text CP}^{(+)}$ projection shown in the left panel of Fig.~\ref{fig:ContributionsImportant}, it becomes less visible compared to other contributions in the Dalitz plot. This analysis highlights that a complete description of CPV phenomenology in this decay requires reproducing both the projected distributions and the Dalitz plot structure.

\begin{figure}[t]
\centering
\includegraphics[width=0.7\linewidth]{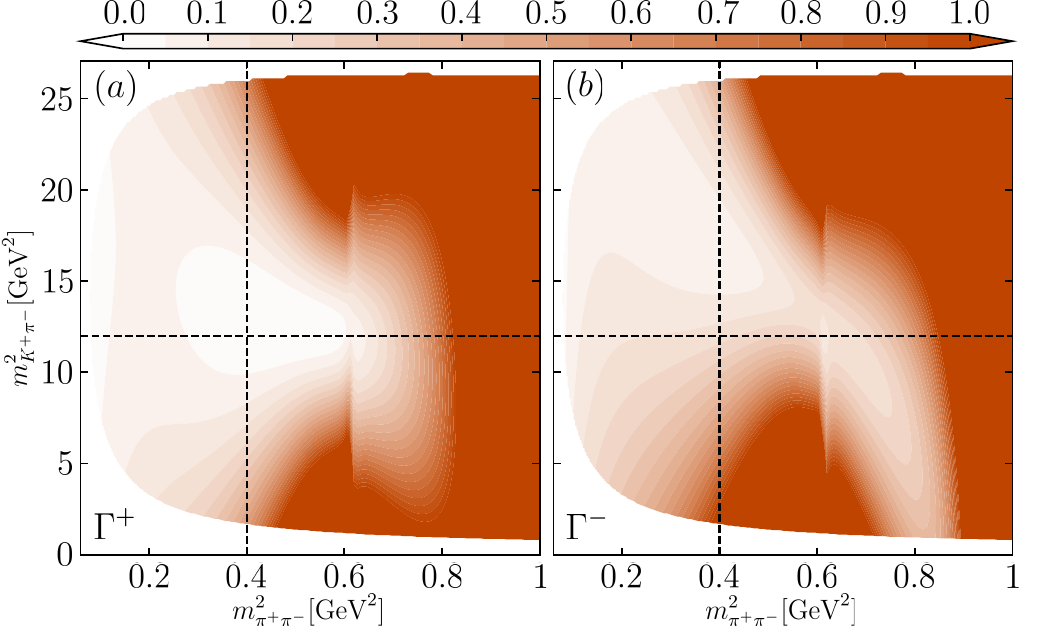}
\caption{
Dalitz plot section, with $s \equiv m^2_{\pi^+\pi^-} \leq 1\GeV^2$, of the decay widths contributing to the CP asymmetry shown in the central panel of Fig.~\ref{fig:Dalitz}. 
Left: $\Gamma^+$, the decay width of $B^+$. 
Right: $\Gamma^-$, the decay width of $B^-$.}
\label{fig:Amps}
\end{figure}

Finally, Fig.~\ref{fig:Amps} displays the Dalitz plot distributions of the decay widths for both $B^+$ and $B^-$. The difference and sum of these distributions define the CP asymmetry through Eq.~(\ref{eq:CPA}). The vertical and horizontal black dashed lines indicate the slices shown in the lower panels of Figs.~\ref{fig:t-section} and \ref{fig:s-section}, respectively, where the $\Gamma^+$ ($\Gamma^-$) slices were shown in red (blue). Note that the $\Gamma^+$ distribution in the left panel of Fig.~\ref{fig:Amps} is strongly suppressed in the region where the large localized CP asymmetry is observed.

\section{Earlier attempts to final-state interactions through the \texorpdfstring{\boldmath{$\mathcal{S}$}}{S} matrix}
\label{app:otherapproaches}

In this appendix, we demonstrate explicitly that, on the $s$-channel unitarity cut, our approach is formally equivalent to that of 
Suzuki and Wolfenstein (SW)~\cite{Suzuki:1999uc,Chua:2007cm}.
Its more practical leading-order expansion in the inelasticity was proposed first in  Ref.~\cite{Wolfenstein:1990ks} (see Ref.~\cite{Bigi:2000yz} for a textbook introduction). In slightly different versions, it is often used in the literature and found application to three-body charmless $B$ meson decays at LHCb~\cite{Bediaga:2013ela, AlvarengaNogueira:2015wpj}. Of course, when expanding the $\mathcal{S}$~matrix, Watson's theorem is only fulfilled approximately.
However, our approach
is more general, and does not call for
approximations, which are not always well justified in practical implementations.

The starting point of both approaches is the relation between the $\mathcal{S}$~matrix and the invariant amplitude $ \Mcal$,
\begin{equation} \label{resonances:eq:M}
  i(2\pi)^4\delta^4 (p_1+p_2-p_{1'}-p_{2'})\,\Mcal(p_1,p_2;p_{1'},p_{2'})_{ba}= _{\text{out}}\!\bra{p_{1'} p_{2'},b} \mathcal{S} - 1 \ket{p_1 p_2,a}_{\text{in}} .
\end{equation}
The unitarity of the $\mathcal{S}$~matrix, represented by the equation $\mathcal{S}^\dagger \mathcal{S} = 1$, is a direct consequence of the conservation of probability. Below the lowest threshold, $\Mcal$  is real. However, once the energy is higher than this threshold, there is a discontinuity associated with the threshold branch point. The  unitarity of the $\mathcal{S}$~matrix relates this discontinuity to the amplitude itself~\cite{Peskin:1995ev}:
\begin{equation} \label{resonances:eq:discM}
  \Disc\, \Mcal_{ba}\nonumber= \Mcal_{ba}-\Mcal^*_{ab}= i \,(2\pi)^4\sum_c \int d\Phi_c \Mcal^*_{cb}\Mcal_{ca} \,,
\end{equation}
where $d\Phi_c$ denotes the invariant phase space for a given channel, labeled as $c$.
For the phase space, we use the conventions provided in Ref.~\cite{ParticleDataGroup:2022pth}
that in particular yield the two-body phase space
\begin{equation} \label{resonances:eq:rhodef}
  \rho_c(s) = \frac{(2\pi)^4}{2} \int \diff\Phi_2 = \frac{1}{16\pi}\frac{2|\vec q_c|}{\sqrt s} \,,
\end{equation}
where $\vec q_c$ denotes the relative momentum of the particle pair in its rest frame.
The unitarity relation for a production amplitude in channel $a$ is represented by
\begin{align}  
  \Disc \, \Acal_a 
  &= \Acal_a-\Acal^{*}_a 
  = i \,(2\pi)^4 \sum_c \int d\Phi_c \Mcal^*_{ca}\Acal_c \,.
   \label{resonances:eq:discA}
\end{align}
To simplify notation, we may now restrict ourselves to $S$~waves. Then the above relations can be written straightforwardly
as
\begin{align}
\mathcal{S} _{ab}&=\delta_{ab}+2i\sqrt{\rho_a} \Mcal_{ab}\sqrt{\rho_b} \,, \label{eq:SvsMsimp}
\\
 \Disc\, \Mcal_{ba}&= 2i\Mcal^*_{cb}\rho_c\Mcal_{ca}  \,, \label{eq:discMsimp}
 \\
 \Disc\, \Acal_{a}&= 2i\Mcal^*_{ba}\rho_b\Acal_{b} \,. \label{eq:discAsimp}
\end{align}
The formalism employed in this work is built on Eq.~\eqref{eq:discAsimp}, 
which in the main text we have rewritten as Eqs.~\eqref{resonances:eq:discAmain} and \eqref{resonances:eq:coupleddiscAmain}, using our notation
tailored for 3-body charmless $B$-meson decays.

In contrast, in Ref.~\cite{Suzuki:1999uc} Suzuki and Wolfenstein (SW) start their derivation of the pertinent production amplitude from
\begin{equation}
\Acal_{a}^{\text{(SW)}}=\mathcal{S}_{ab}\Acal_{b}^{\text{(SW)}\, *} \,,
\label{eq:SWstart}
\end{equation}  
which has the formal solution
\begin{equation}
    \Acal_{a}^{\text{(SW)}} =
\mathcal{S}_{ab}^{1/2}\Acal_{b\, 0}^{\text{(SW)}} \ ,
\label{eq:SWsol}
\end{equation}
where $\Acal_{b\, 0}^{\text{(SW)}}$ is a vector in channel space, which is real in the absence of $T$ violation. 
In this Appendix, we provide  the relation between the production amplitude employed in this work and
that of Ref.~\cite{Suzuki:1999uc}.
To derive the discontinuity of $\Acal_{a}^{\text{(SW)}}$ we subtract $\Acal_{a}^{\text{(SW)}\, *}$ from Eq.~\eqref{eq:SWstart}
and employ Eq.~\eqref{eq:SvsMsimp}: 
\begin{equation}
 \Disc\, \Acal_{a}^{\text{(SW)}} = 2i\sqrt{\rho_a} \Mcal_{ab}\sqrt{\rho_b}\, \Acal_{b}^{\text{(SW)}\, *} \,.
 \label{eq:discASWsimp_a}
\end{equation}
Taking the complex conjugate and transpose on both sides, this reads
\begin{equation}
  \Disc\, \Acal_{a}^{\text{(SW)}} = 2i\sqrt{\rho_a} \Mcal_{ba}^*\sqrt{\rho_b}\, \Acal_{b}^{\text{(SW)}} \,.
 \label{eq:discASWsimp_b}
\end{equation}
We may now introduce 
\begin{equation}
\sqrt{\rho_a} \Acal_a =  \Acal_{a}^{\text{(SW)}} 
\label{eq:AsvASW}
\end{equation}
to obtain
\begin{equation}
 \Disc\, \left\{\sqrt{\rho_a}  \Acal_a\right\} = 2i\sqrt{\rho_a} \Mcal_{ba}^*{\rho_b}\,   \Acal_b\,.
 \label{eq:discASWsimp2}
\end{equation}
Once we factorize out the phase-space factor  
from the discontinuity, 
we recover Eq.~\eqref{eq:discAsimp},  which, as commented above, is our starting point. Thus, formally, both the SW  Eq.~\eqref{eq:SWsol} and our approach solve the same constraint over the $s$-channel discontinuity. 
However, note that Eq.~\eqref{eq:discAsimp} does not fix, for instance, the modulus of  the solution. Actually, we can multiply $\Acal_a$ on the left by any real analytic matrix function $A(s)$ on the real axis above the lowest threshold and still obtain a solution. Then, it is a simple exercise to show that, in matrix form,  $\Omega(s)=\mathcal{S}^{1/2}(s)A(s)$.

However, there are several advantages in following the Omn\`es approach over the SW method.
First, $\Acal_{a}^{\text{(SW)}}$ inherits the analytic properties of $\mathcal{S}^{1/2}_{ab}$, some of which are spurious for $\Acal_{a}^{\text{(SW)}}$; for instance, additional (left) cuts present in $\mathcal{S}_{ab}$, but not in $\Omega_{ab}$.  Moreover, for a given $\mathcal{S}_{ab}$, its square root is not uniquely defined. Implicit in the SW derivation, the choice of square-root must satisfy $\mathcal{S}^{1/2}=\mathcal{S}(\mathcal{S}^{1/2})^*$.\footnote{Formally again, this can be obtained from the diagonal form of $\mathcal{S}$ by an orthogonal (not unitary) transformation, which can be easily shown to exist.}

Let us also recall that our aim is to factor out the FSI or rescattering effects, which is not necessarily well accomplished by the SW method. This is easy to realize when the $\mathcal{S}$~matrix is elastic in a given FSI channel, which is the case of interest in the kinematic regime studied in this work. Then, Eq.~\eqref{eq:SWsol} becomes
$\Acal^{\text{(SW)}}=\mathcal{S}^{1/2}\Acal_{0}^{\text{(SW)}}$, but in the elastic case $\mathcal{S}^{1/2}(s)=e^{i\delta(s)}$, with $\delta(s)$ the scattering phase shift in that channel. As expected, $\Acal^{\text{(SW)}}$ fulfills Watson's theorem. However, $\vert \Acal^{\text{(SW)}}\vert^2 =\vert \Acal_{0}^{\text{(SW)}}\vert^2$, which means that to describe a resonant shape in a term proportional to that squared amplitude, it has to be incorporated in the real $\Acal_{0}^{\text{(SW)}}$, which was supposed to describe the source, not rescattering. As we have discussed in Sec.~\ref{sec:applicationtoB3M}, such  contributions proportional to  $\vert \Acal\vert^2$ not only exist, but in our case play an essential role to describe the $\rho$ line shape in the CP asymmetry. Within our formalism, we factorize the FSI in the Omn\`es functions, whose moduli $\vert \Omega\vert$ are independent of the source and capture the resonant shapes when present, as seen in Fig.~\ref{fig:omnes_functions}. 
However, in many other implementations of FSI, resonant shapes are also included in the ``source" terms of the formalism. This lack of a complete FSI factorization is then propagated to any other interference contribution containing such a resonant source term. A relatively similar situation occurs in the coupled-channel case with the SW formalism, although it is not so evidently noticed.

Moreover, as already commented, given the difficulty in calculating $\mathcal{S}^{1/2}$, in practical applications, Eq.~\eqref{eq:SWsol}
is employed in conjunction with some Taylor expansion of
the $\mathcal{S}$~matrix. For instance,  Eq.~\eqref{eq:SvsMsimp} is frequently rewritten in terms of the $t$~matrix as $\mathcal{S}=\mathbf{1}+2it$, and the following expansion is considered: $\mathcal{S}^{1/2}=\mathbf{1}+it+\ldots$~\cite{Bediaga:2013ela,AlvarengaNogueira:2015wpj,Cheng:2016shb,Garrote:2022uub}.\footnote{Beware of the factor of 2 in the definition of the $t$~matrix, which customarily appears in the  partial-wave basis. Otherwise one may confuse the $\mathcal{S}$ and $\mathcal{S}^{1/2}$ expansions (see the discussion in Ref.~\cite{Garrote:2022uub}).} In the popular Wolfenstein model~\cite{Wolfenstein:1990ks,Bigi:2000yz}, it is assumed that 
$\mathcal{S}=\mathcal{S}_0+\mathcal{S}_1$, with $\mathcal{S}_0$ unitary and diagonal and $\mathcal{S}_1$ a small correction treated perturbatively to first order.
Both treatments can be recast as a leading-order expansion in the inelasticity or a related parameter, which may not be well justified and may lead to large violation of Watson's theorem. Such approximations are not necessary in our approach.

While $\mathcal{S}^{1/2}$ can be evaluated explicitly under strongly simplifying assumptions—for instance, by imposing $\delta_{K\bar K\to K\bar K}=\delta_{\pi\pi\to\pi\pi}=\delta_{\pi\pi\to K\bar K}/2$~\cite{Cheng:2016shb,Cheng:2020ipp}—this remains a crude approximation. Nonetheless, it is frequently adopted in the literature for the $S0$-wave~\cite{Bediaga:2013ela,AlvarengaNogueira:2015wpj,Cheng:2007si,Cheng:2020ipp}, including several LHCb analyses~\cite{Aaij:2019qps,Aaij:2019jaq,Aaij:2019hzr}. In practice, this ansatz fails to accurately describe the existing data for $\delta_{\pi\pi\to\pi\pi}$ and $\delta_{\pi\pi\to K\bar K}$ (see the discussion in Ref.~\cite{Garrote:2022uub}). In contrast, our Omn\`es functions are determined directly from data-driven dispersive analyses of $\pi\pi$ and $\pi\pi\to K\bar K$ scattering~\cite{Garcia-Martin:2011iqs,Pelaez:2018qny,Pelaez:2020gnd}.

\bibliographystyle{jhep_mod}
\bibliography{CPV_refs}

\end{document}